\documentclass[a4paper]{llncs}
\usepackage{amssymb,latexsym,amsxtra,amscd,amsfonts}
\begin{document}
\pagestyle{plain}


\newcommand{\cO}{{\mathcal O}}
\newcommand{\cE}{{\mathcal E}}
\newcommand{\cI}{{\mathcal I}}
\newcommand{\cF}{{\mathcal F}}
\newcommand{\cR}{{\mathcal R}}

\newcommand{\Z}{\mathbb Z}
\newcommand{\N}{\mathcal N}
\newcommand{\cS}{\mathcal S}
\newcommand{\cL}{{\mathcal L}}
\newcommand{\F}{\mathbb F}
\newcommand{\C}{\mathbb C}
\newcommand{\D}{\mathbb D}
\newcommand{\Q}{\mathbb  Q}
\newcommand{\G}{\mathbb G}
\newcommand{\m}{\mathfrak m}
\newcommand{\n}{\mathfrak n}
\newcommand{\f}{\mathfrak f}
\newcommand{\e}{\mathfrak e}
\newcommand{\g}{\mathfrak g}
\newcommand{\h}{\mathfrak h}
\newcommand{\be}{\mathfrak b}

\newcommand{\q}{\mathfrak q}
\newcommand{\pe}{\mathfrak p}
\newcommand{\ma}{\mathfrak a}

\newcommand{\mF}{\mathcal F}
\newcommand{\mG}{\mathcal G}

\newcommand{\tF}{\tilde{\mathcal F}}
\newcommand{\tG}{\tilde{G}}
\newcommand{\tx}{\tilde{x}}
\newcommand{\td}{\tilde{d}}
\newcommand{\tf}{\tilde{f}}
\newcommand{\ta}{\tilde{a}}
\newcommand{\te}{\tilde{e}}
\newcommand{\tM}{\tilde{M}}
\newcommand{\txF}{\widetilde{x_iF}}
\newcommand{\tI}{\tilde{I}}
\newcommand{\tu}{\tilde{u}}

\newcommand{\Fp}{\F_p}
\newcommand{\Fq}{\F_q}

\newcommand{\ka}{{\kappa}}
\newcommand{\reg}{\rm reg}
\newcommand{\End}{\rm End}
\newcommand{\Pic}{\rm Pic}
\newcommand{\NS}{\rm NS}
\newcommand{\Div}{\rm div}
\newcommand{\nuinf}{\nu_{\infty}}
\newcommand{\siminf}{\sim_{\infty}}
\newcommand{\bu}{{\bf u}}
\newcommand{\bt}{{\bf \theta}}
\newcommand{\btt}{{\bf t}}
\newcommand{\bz}{{\bf z}}
\newcommand{\bx}{{\bf x}}
\newcommand{\bX}{{\bf X}}
\newcommand{\by}{{\bf y}}
\newcommand{\bg}{{\bf g}}

\newcommand{\la}{\langle}
\newcommand{\ra}{\rangle}

\title{Algebraic blinding and cryptographic trilinear maps}
\author{Ming-Deh A. Huang (USC, mdhuang@usc.edu)}
\institute{Computer Science Department,University of Southern California, U.S.A.}


\maketitle

\begin{abstract}
It has been shown recently that cryptographic trilinear maps are  sufficient for achieving indistinguishability obfuscation. In this paper we develop algebraic blinding techniques for constructing such maps.
An earlier approach involving Weil restriction can be regarded as a special case of blinding in our framework. However, the techniques developed in this paper are more general, more robust, and easier to analyze.
The trilinear maps constructed in this paper are efficiently computable.
The relationship between the published entities and the hidden entities under the blinding scheme is described by algebraic conditions. Finding points on an algebraic set defined by such conditions for the purpose of unblinding is difficult as these algebraic sets have dimension at least linear in $n$ and involves  $\Omega(n^2)$ variables, where $n$ is the security parameter. Finding points on such algebraic sets in general takes time exponential in $n^2\log n$ with the best known methods. Additionally these algebraic sets are characterized as being {\em triply confusing} and most likely {\em uniformly confusing} as well. These properties provide additional evidence that efficient algorithms to find points on such algebraic sets seems unlikely to exist.
In addition to algebraic blinding, the security of the trilinear maps also depends on the computational complexity of a trapdoor discrete logarithm problem which is defined in terms of an associative non-commutative polynomial algebra acting on torsion points of a blinded product of elliptic curves.
\end{abstract}
\section{Introduction}
In this paper we develop algebraic blinding techniques for the construction of cryptographically interesting trilinear maps. Cryptographic applications of $n$-multilinear maps for $n >2$ were first proposed in the work of  Boneh and Silverberg \cite{BS}.  However the existence of such maps remains an open problem \cite{BS,HR}. The problem has attracted much attention recently as multilinear maps and their variants prove to be useful for indistinguishability obfuscation. More recently Lin and Tessaro \cite{LT} showed that trilinear maps are sufficient for the purpose of achieving indistinguishability obfuscation (see \cite{LT} for references to related works along several lines of investigation).
This striking result brought the following question into the spotlight: can a cryptographically interesting trilinear map be constructed?

\ \\The results of this paper follow a line of investigation initiated by an observation of Chinburg (at the AIM workshop on cryptographic multilinear maps (2017)) that the following map from \'{e}tale cohomology may serve as the basis of constructing a cryptographically interesting trilinear map:

\[ H^1 (A,\mu_{\ell})\times H^1 (A,\mu_{\ell})\times H^2 (A,\mu_{\ell})\to H^4 (A, \mu_{\ell}^\otimes{3})\cong \mu_{\ell}\]
where $A$ is an abelian surface over a finite field $\F$ and the prime $\ell\neq {\rm char}(\F)$.
Following up on Chinburg's idea, a method for constructing trilinear maps was proposed in \cite{H1,H2}.  It was based on the following map that can be derived from the cohomological map just mentioned:  $(\alpha,\beta,\cL) \to e_{\ell}(\alpha,\varphi_{\cL} (\beta))$, where  $\alpha,\beta\in A[\ell]$, $\cL$ is an invertible sheaf, and $\varphi_{\cL}$ is the map $A\to A^* =\Pic^0 (A)$ so that
\[ \varphi_{\cL} (a) =t_a^*\cL \otimes \cL^{-1} \in \Pic^0 (A)\]
for $a\in A(\bar{\F})$ where $t_a$ is the translation map defined by by $a$ (\cite{MilneA} \S~1 and \S~6).
In the map just described one no longer needs to assume that $A$ is of dimension 2, and the third participant $\cL$  in the trilinear map can be identified with an endomorphism of $A$.  With this approach the third group in the pairing is to be constructed from endomorphisms of $A$, and the challenge is to encode the endomorphisms involved in such a way that the resulting group has hard discrete logarithm problem.  The method proposed in \cite{H3} tackles this issue by using Weil descent (or Weil restriction) \cite{DG,F,FL,W}.
The trilinear map in \cite{H3} is derived from a blinded version of the following trilinear map:
\[
\begin{array}{rcl}
A[\ell]^d \times A[\ell]^d \times Mat_d (\F_{\ell}) & \to & \mu_{\ell}\\
(\alpha,\beta,M) &\to & e(\alpha, M(\beta))
\end{array}
\]
where $\alpha,\beta\in A[\ell]^d$, $M\in Mat_d(\F_{\ell})\subset \End (A[\ell]^d)$, and $e$ is a non-degenerate bilinear pairing on $A[\ell]^d$ (determined by a non-degenerate bilinear pairing on $A[\ell]$).  The blinding of the map just described involves Weil descent.

\ \\In this paper we develop algebraic blinding techniques for constructing trilinear maps.  The blinding in \cite{H3} involving Weil descent can be regarded as a special case in our framework.  In comparison, blinding using Weil descent is more restrictive, and the analysis is more complicated.  The blinding techniques developed in this paper are more robust, more general, and easier to analyze. The trilinear maps constructed in this paper are efficiently computable. Under our algebraic blinding system, the relationship between the published entities and the hidden entities is described by algebraic conditions.  Finding a point on an algebraic set defined by such conditions implies uncovering the blinding at least partially up to local isomorphism.  However Theorem~\ref{specify-trilinear} shows that such an algebraic set has dimension at least linear in $n$ and involves  $\Omega(n^2)$ variables, where $n$ is the security parameter. Solving such non-linear polynomial systems in general takes expected time $2^{O(n^2\log n)}$ if the polynomials are of bounded degree (\cite{Ay,HW}).
Theorem~\ref{specify-trilinear} also shows that these algebraic sets are {\em triply confusing} and most likely {\em uniformly confusing}.
These properties, as defined and discussed \S~\ref{semi-local-functions}, provide additional evidence that efficient algorithms to find points on such algebraic sets seems unlikely to exist.

\ \\In addition to algebraic blinding, the security of the trilinear maps constructed in this paper also depends on the computational complexity of a trapdoor discrete logarithm problem presented in \S~\ref{trapdoor-dl}.   The problem is defined in terms of an associative non-commutative polynomial algebra acting on torsion points of a blinded product of elliptic curves. The kernel ideal of such action is hidden due to blinding, except polynomially many elements in the kernel are made public.

\ \\In our construction the blinding parameters are secretly chosen and the elements of the third group in the pairing require private encoding.  It remains an interesting open problem whether a trilinear map without private encoding, perhaps along the line of Chinburg's idea or the approach in \cite{H1,H2}, can be constructed.

\section{Algebraic blinding systems and trilinear map construction}
The blinding scheme developed in this paper can be easily adapted to a broader context.  However, to simplify presentation we restrict ourselves to the situation where the objects to be blinded are points and maps on $\prod_{i=1}^n V_i (K)$ where $V_i$ is an
algebraic set contained in $\bar{k}^2$ defined over a finite field $K$ over a smaller finite field $k=\F_q$. For the trilinear map construction $V_i$ will be an elliptic curve isomorphic over $K$ to some elliptic curve $E$ chosen from an isogeny class of pairing friendly elliptic curves, and $n$ is linear in the security parameter.

\ \\In this restrictive framework, a basic blinding map $\rho$ is an isomorphism over $K$ from a subset $W$ of $\bar{k}^{3n}$ onto $\bar{k}^{2n}$.   Let $\hat{V}$ be the inverse image of $\prod_{i=1}^n V_i$ under  $\rho$.  We say that $\prod_{i=1}^n V_i$ is blinded in $\hat{V}$ by the secret map $\rho$. We say that $(\alpha_i)_{i=1}^n\in \prod_{i=1}^n V_i (K)$ blinded in $\hat{\alpha}\in\hat{V}(K)^n$ if $\rho(\hat{\alpha}) = (\alpha_i)_{i=1}^n$.

\ \\The maps to be blinded are local in nature. For example, a map that arises in the trilinear map construction is of the form $\varphi=(\varphi_i)_{i=1}^n:\prod_{i=1}^n V_i \to \prod_{i=1}^n V_i$ where $\varphi_i:\prod_{j=1}^n V_j \to V_i$ is of locality two in the sense that $\varphi_i ((\alpha_j)_{j=1}^n)$ depends on $\alpha_{i_1}$ and $\alpha_{i_2}$ for some $i_1$ and $i_2$.   More specifically, $V_i$ is isomorphic to an elliptic curve $E$ for all $i$, $\varphi_i ((\alpha_j)_{j=1}^n)$ is the point on $V_i\simeq E$ corresponding to the sum of the points $\alpha_{i_1}\in V_{i_1}\simeq E$ and $\alpha_{i_2}\in V_{i_2}\simeq E$.   Thus $\varphi$ determines an element of $\End (E^n [\ell])$  when it is restricted to $\prod_{i=1}^n V_i [\ell]$, and can be identified with $M\in Mat_n (\F_{\ell})\subset \End (E^n [\ell])$, where the $i$-th row of $M$ is all 0s except two 1s at the entries $(i,i_1)$ and $(i,i_2)$ .

\ \\The secret map $\varphi$, hence the secret matrix $M$, is blinded in $\hat{\varphi} = \rho^{-1}\varphi\rho: \hat{V}\to\hat{V}$.  Suppose $\rho(\hat{\alpha})=(\alpha_i)_{i=1}^n$, $\rho(\hat{\beta})=(\beta_i)_{i=1}^n$ and $\hat{\varphi} (\hat{\alpha}) = \hat{\beta}$.   Then $(\hat{\alpha}, \hat{\beta})$ hides the pair $(\alpha_i)_{i=1}^n$ and $(\beta_i)_{i=1}^n$ and the fact that $\beta_i = \varphi_i (\alpha_{i_1},\alpha_{i_2})$.

\ \\We make some remark about the security of blinding.  Suppose $\hat{\varphi}$ is specified and made accessible to the public.  If $\rho$ can be  efficiently discovered say from evaluation of $\hat{\varphi}$ at publicly available sampled points of $\hat{V}$, then the blinding is certainly insecure since $\varphi$ is revealed as
$\varphi = \rho\hat{\varphi}\rho^{-1}$.  More generally we consider the blinding compromised if some $\rho':W\to\bar{k}^{2n}$, and $\varphi'=(\varphi_i)_{i=1}^n : \prod_{i=1}^n V'_i \to \prod_{i=1}^n V'_i$ of locality 2 can be efficiently constructed in the public such that $\rho'$ maps $\hat{V}$ isomorphically onto $\prod_{i=1}^n V'_i$, and ${\rho'}^{-1}\varphi'\rho'=\hat{\varphi}$.    Let $\rho'=(\rho'_i)_{i=1}^n$ where $\rho'_i : \bar{k}^{3n}\to \bar{k}^2$. If ${\rho'}^{-1}\varphi'\rho'=\hat{\varphi}$ then $\varphi'_i \circ (\rho'_{i_1} , \rho'_{i_2})=\rho'_i \hat{\varphi}$ for $i=1,\ldots,n$. We also consider the blinding compromised if for some $i$, $\rho'_i$, $\rho'_{i_1}$ and $\rho'_{i_2}$ and $\varphi'_i$ are found such that $\varphi'_i \circ (\rho'_{i_1} , \rho'_{i_2})=\rho'_i \hat{\varphi}$.  The left hand side of the equality,   $\varphi'_i \circ (\rho'_{i_1} , \rho'_{i_2})$, is an example of {\em semi-local decomposition} studied in \S~\ref{semi-local-functions}.

\subsection{Blinding system}
\ \\Our blinding system involves maps from general linear groups, quadratic isomorphisms of affine spaces of small dimension, and Frobenius twists.

\ \\{\bf Maps from general linear groups} Let $Gl_m (K)$ be the general linear group over $K$ where
each $A\in Gl_m (K)$ can be identified with an $m$ by $m$ invertible matrices $(a_{ij})$ with $a_{ij}\in K$ for $1\le i,j \le m$, so that for $\bx=(x_1,\ldots,x_m)\in\bar{k}^m$, $A(\bx) = (\sum_{j=1}^m a_{ij}x_j)_{i=1}^m$.

 \ \\Suppose the algebraic set to be blinded is the product of $n$
algebraic sets contained in $\bar{k}^2$.  Then the general linear groups involved are $Gl_m (K)$ where $m=3n$ and $m=2, 3$.


\ \\{\bf Local quadratic isomorphisms} Let $p(x)\in K[x]$, $q(x,y)\in K[x,y]$ where $\deg p(x)=\deg q(x,y)=2$.  Then $(x,y,z)\to (x, y+p(x), z+q(x,y)$ defines an isomorphism $\bar{k}^3 \to \bar{k}^3$, denoted as $\lambda_{p,q}$.

\ \\Let $A,B\in Gl_3 (K)$ and let $p(x)\in K[x]$, $q(x,y)\in K[x,y]$ where $\deg p(x)=\deg q(x,y)=2$.  Then $\lambda=B\circ\lambda_{p,q}\circ A$ defines an isomorphism $\bar{k}^3 \to \bar{k}^3$.  For $\bx\in\bar{k}^3$, $\lambda(\bx) = (f_i(\bx))_{i=1}^3$ where $f_i$ is a quadratic polynomial in $K[x,y,z]$ for $i=1,2,3$.  For random choices of $p,q,A,B$, the $f_i$'s are most likely dense.

\ \\{\bf Blinding space and blinding maps}  Let $\lambda=B\circ\lambda_{p,q}\circ A$ be as discussed above with $A,B\in Gl_3 (K)$ and  $p(x)\in K[x]$, $q(x,y)\in K[x,y]$ where $\deg p(x)=\deg q(x,y)=2$.  Let $\mu=\lambda^{-1}$. Let $\tilde{\mu}$ be the map $\bar{k}^2\to\bar{k}^3$ such that $\tilde{\mu} (x,y) = \mu (x,y,y)$, and let $W_{\lambda}$ be the isomorphic image of $\bar{k}^2$ under $\tilde{\mu}$.  Then $W_{\lambda}$ is isomorphic to $\bar{k}^2$.

\ \\We form a blinding space $W\subset \bar{k}^{3n}$, where $W$ is isomorphic to $\bar{k}^{2n}$, as follows. For $i=1,\ldots,n$, choose a random $\lambda_i=B_i\circ\lambda_{p_i,q_i}\circ A_i$ in the manner as discussed above with $A_i,B_i\in Gl_3 (K)$ and  $p_i(x)\in K[x]$, $q_i(x,y)\in K[x,y]$ where $\deg p_i(x)=\deg q_i(x,y)=2$.  Let $W_i = W_{\lambda_i}$.  Thus $W_i$ is the zero set of $f_{i2}-f_{i3}$ where $\lambda_i (\bx) = (f_{ij} (\bx))_{j=1}^3$ where $\bx = (x,y,z)$.
Choose a random $\delta\in Gl_{3n} (K)$.  Let $W= \delta^{-1} \prod_{i=1}^n W_i$.

\ \\Write $\delta = (\delta_i)_{i=1}^n: \bar{k}^{3n}\to \prod_{i=1}^n \bar{k}^3$ with $\delta_i:\bar{k}^{3n}\to \bar{k}^3$ given by linear forms $L_{ij}$, $j=1,2,3$, in $3n$ variables.   For $i=1,\ldots,n$,
$\lambda_i \circ \delta_i = (F_{ij})_{j=1}^3$ where  $F_{ij}=f_{ij}\circ (L_{i1}, L_{i2}, L_{i3})$ for $j=1,2,3$.
Let $\rho_i =pr\circ\lambda_i\circ\delta_i = (F_{i1}, F_{i2})$ where $pr$ denotes the projection $\bar{k}^3\to \bar{k}^2:(x,y,z)\to (x,y)$.
Then $W_i = \delta_i (W)$ and $W$ is the zero set of
$\{ F_{i2}-F_{i3} : i=1,\ldots,n\}$.

\ \\The basic blinding map associated with the blinding space $W$ is $\rho: \bar{k}^{3n} \to \prod_{i=1}^n \bar{k}^2$ where $\rho=(\rho_i)_{i=1}^n$.  We see that $\rho$ maps $W$ isomorphically to $\bar{k}^{2n}$.

\ \\{\bf Frobenius twists}  Suppose $[K:k]=d$.  For simplicity we assume $d=O(n)$. Let $\tau$ denote the Frobenius map $x\to x^q$ for $x\in\bar{k}$.  Let $\tau_a$ also denote  $\tau^a$ for $a\in\Z$.
For $0\le a,b\le d-1$, let $\tau_{a,b}$ denote the map $\tau_{a,b}:\bar{k}^2\to\bar{k}^2$ such that $\tau_{a,b}(x,y)=(\tau_a (x), \tau_b (y))=(x^{q^a},y^{q^b})$ for $x,y\in\bar{k}$.

\ \\A blinding map on $W$ is the basic blinding map twisted by Frobenius locally, that is, $(\tau_{a_i,b_i}\circ\rho_i)_{i=1}^n$ with $0\le a_i,b_i \le d-1$ for $i=1,\ldots,n$.

\ \\{\bf Weil descent as a special case}  Suppose $V\subset\bar{k}^2$ is an elliptic curve defined over $K$.  Suppose $[K:k]=d$ as above.  Then a Weil descent of $V$ from $K$ to $k$ can be identified with $\hat{E}=\delta^{-1} \prod_{i=1}^d V_i$ where
$V_i = V^{\tau_i}$ for $i=0,\ldots,d-1$, where $\delta\in Gl_{2d} (K)$ is determined by a basis $\bt$ of $K$ over $k$ as follows.  Organize the coordinates of $\bar{k}^{2d}$ in two vectors $\hat{x}= x_0,\ldots,x_{d-1}$ and $\hat{y} = y_0,\ldots,y_{d-1}$.  For $\alpha = (\alpha_i)_{i=0}^{d-1}$ and $\beta = (\beta_i)_{i=0}^{d-1} \in \bar{k}^d$, let
$\la \alpha,\beta \ra = \sum_{i=0}^{d-1} \alpha_i \beta_i$. Then $\delta= (\delta_i)_{i=0}^{d-1}$ where
\[ \delta_i  (\hat{x},\hat{y}) = (\la\hat{x} , \bt^{\tau_i} \ra, \la\hat{y} ,\bt^{\tau_i} \ra)\in V_i \]
for $i=0,\ldots,d-1$.  Note that $\delta$ is determined by the matrix in $Gl_d (K)$ with $\bt^{\tau_i}$ as the $i$-th row for $i=0,\ldots,d-1$.  The analysis in \cite{H3} shows that that Weil descent alone is not sufficient for the purpose of blinding, as a result additional local birational maps are involved in the trilinear map construction there.  However the restrictive nature of $\delta$, especially the fact that $\delta$ is determined by a $K/k$-basis, makes things complicated both in terms of construction and analysis.  In contrast the techniques developed in this paper are more general and more robust at the same time.

\ \\In this paper we will focus on basic blinding maps for the most part since they are sufficient for our purposes.   We remark that adding Frobenius twists to basic blinding maps provides an additional layer of protection and allows us to pay attention to the fact that the functions and maps of interest are applied to $K$-rational points.  We will discuss blinding maps with Frobenius twists in \S~\ref{Frob}. The readers may skip \S~\ref{Frob} for the first reading and assume that the blinding map involved is a basic blinding map.

\subsection{Trilinear map construction}
\ \\To construct a trilinear map we start by choosing from the isogeny class of a pairing friendly elliptic curve some $E/K$ defined by $y^2=x^3+ax+b$ with $a,b\in K$.  Suppose $E[\ell] \subset E(K)$, and $\log\ell$ and $\log |K|$ are linear in the security parameter $n$.  In case Frobenius twists are involved in forming the blinding map then we assume $K$ is a finite extension over a finite field $k=\F_q$,
$[K:k]=O(n)$, and a basis of $K/k$ is published.  The curve $E$ is considered secret.

\ \\Our trilinear map is derived from a blinded version of the following map:
\[
\begin{array}{rcl}
E[\ell]^n \times E[\ell]^n \times Mat_n (\F_{\ell}) & \to & \mu_{\ell}\\
(\alpha,\beta,M) &\to & e(\alpha, M(\beta))
\end{array}
\]
where $\alpha,\beta\in E[\ell]^d$, $M\in Mat_n (\F_{\ell})$ is identified with an element of $\End (E[\ell]^n)$, and $e$ is a non-degenerate bilinear pairing on $E[\ell]^n$ naturally induced by Weil pairing on $E[\ell]$.

\ \\For $A\in Gl_2 (K)$, let $E^A$ denote the elliptic curve which is the image of $E$ under $A$.  Let $\overline{E^A}$ denote the image of $E^A$ under $\jmath:(x,y)\to (x^{-1}, y^{-1})$. Consider the set $\{ \overline{E^A}:
A=\left( \begin{array}{cc} a & b \\ c & d \end{array}\right)\in Gl_2 (K)$, neither $a=d=0$ nor $b=c=0\}$.  Choose randomly from this family $E_i$, $i=1,\ldots,n$.  The reason for considering the set of $\overline{E^A}$ will be made clear in \S~\ref{semi-local-functions}.

\ \\Choose a random blinding map $\rho=(\rho_i)_{i=1}^n$.
Let $\hat{E}=\rho^{-1} \prod_{i=1}^n E_i$.
Choose $\alpha,\beta\in E[\ell]^n$ such that $e(\alpha,\beta)\neq 1$. Note that $e$ is the natural extension of Weil pairing $e_E$ on $E$, so that for $\alpha=(\alpha_i)_{i=1}^n$ and $\beta=(\beta_i)_{i=1}^n$ in $E[\ell]^n$ with $\alpha_i,\beta_i\in E[\ell]$ for $i=1,\ldots,n$,
$e(\alpha,\beta) =\prod_{i=1}^n e_E (\alpha_i,\beta_i)$.  Let $\hat{\alpha},\hat{\beta}\in \hat{E}[\ell]$ such that $\hat{\alpha}$ corresponds to $\alpha$ and $\hat{\beta}$ corresponds to $\beta$ under
$\hat{E}\stackrel{\rho}{\to} \prod_{i=1}^n E_i \simeq E^n$.
Let $G_1$ and $G_2$ be the groups generated by
$\hat{\alpha}$ and $\hat{\beta}$ respectively.  They are the first two groups in the trilinear map.  The points $\hat{\alpha}$ and $\hat{\beta}$ are made public, while $\alpha$ and $\beta$ are secret.
The addition map on $\hat{E}[\ell]$, $\hat{m}$, serves as the group law in both $G_1$ and $G_2$.

\ \\Choose a set of $N=O(n^2)$ matrices $M_i \in Gl_n (\F_{\ell})$ such that each row of $M_i$ has two non-zero entries, which contain 1, and that the matrices $M_i$ together with the identity matrix $M_0$ generate $Mat_n (\F_{\ell})$ as a vector space over $\F_{\ell}$.  Associate $M_i$ with the endomorphism $\varphi_i\in \End \prod_{i=1}^n E_i \simeq \End E^n$ as follows.  Suppose $\alpha\in\prod_{i=1}^n E_i  [\ell]$ is identified with
$(\alpha_i)_{i=1}^n$, with $\alpha_i \in E[\ell]$,
under $\prod_{i=1}^n E_i [\ell] \simeq E^n[\ell]$.   If $(j,j_1)$ and $(j,j_2)$ are the two non-zero entries of the $j$-th row of $M_i$, then
$\varphi_i (\alpha)$ is identified with $(\beta_j)_{j=1}^n$ where $\beta_j=m (\alpha_{j_1}, \alpha_{j_2})$.  Let $\hat{\varphi}_i = \rho^{-1}\circ\varphi_i\circ\rho$ for $i=1,\ldots, N$.

\ \\Let $R=\F_{\ell} [z_1,\ldots,z_N]$ be an associative non-commutative $\F_{\ell}$-algebra generated by variables $z_1,\ldots,z_N$.  Define an action of $R$ on $\hat{E}[\ell]$ so that $z_i$ acts by $\hat{\varphi}_i$ for $i=1,\ldots,N$.  This is compatible with the action of $R$ on $E^n [\ell]$ where $z_i$ acts by $M_i$. Let $\Lambda$ be the kernel of the $\F_{\ell}$-algebra morphism $\lambda: R\to Mat_n (\F_{\ell})$ determined by
$\lambda (z_i ) = M_i$, $i=1,\ldots, N$.

\ \\We are ready to describe the trilinear map: $G_1\times G_2\times G_3 \to \mu_{\ell}\subset K$.

\ \\The groups $G_1$ and $G_2$ are generated respectively by some $\hat{\alpha},\hat{\beta}\in\hat{E}[\ell]$ with
$\hat{e} (\hat{\alpha},\hat{\beta} ) \neq 1$.

\ \\The group $G_3 = \F_{\ell}$ is identified with $(\F_{\ell}+\Lambda)/\Lambda$.
 In general $a\in\F_{\ell}$ can be represented by polynomials in $a+\Lambda$.  However for efficiency purpose we will only choose polynomials in $a+\Lambda$ of degree $n^{O(1)}$ with number of terms with nonzero coefficients bounded in $n^{O(1)}$.  For simplicity let $[a]$ denote the subset of $f\in a+\Lambda$ such that $f$ is a linear polynomial plus a term of degree $n$.  We can allow more general $f\in a+\Lambda$ to be included in $[a]$ as long as the support of $f$ is polynomially bounded in $n$.  The choice just made is simple but sufficient for our purposes.  More explicitly we use the following procedure to encode $a\in\F_{\ell}$.

\ \\{\bf Private encoding} To encode $a\in\F_{\ell}$, choose random $i_1,\ldots,i_n\in \{1,\ldots,N\}$, then find $c$ and $b_0,\ldots,b_N\in\F_{\ell}$ such that $c M_{i_1}\ldots M_{i_n} + \sum_{i=0}^N b_i M_i = a$.  Then set $f = c z_{i_1}\ldots z_{i_n} + \sum_{i=0}^N b_i z_i$.   We have $f\in [a]$.
Note that $c$ and $b_i$ can be found by simple linear algebra once $M=M_{i_1}\ldots M_{i_n}$ is computed.

\ \\For $a,b,c\in\F_{\ell}$ and $f\in [c]$, the trilinear map sends
$(a\hat{\alpha},b\hat{\beta},f)$ to $\hat{e} (a\hat{\alpha}, f(b\hat{\beta}))=\zeta^{abc}$, where
$\zeta = \hat{e} (\hat{\alpha},\hat{\beta})$.  Note that for $\gamma\in \hat{E}[\ell]$,
$f(\gamma) = c \hat{\varphi}_{i_1}\ldots \hat{\varphi}_{i_n} (\gamma) + \sum_{i=0}^N b_i \hat{\varphi}_i (\gamma)$.

\ \\For the computation of $\hat{e}$ two functions $\hat{g}$ and
$\hat{h}$ are specified, both are products of semi-local functions of bounded degree, as will be discussed in \S~\ref{pairing}.

\subsection{Security of trilinear map}
To summarize the discussion up to this point, in constructing the trilinear map a secret random blinding map $\rho$ is applied to $\prod_{i=1}^n E_i$ where each $E_i$ is of the form $\overline{E^A}$ with secret random $A=\left( \begin{array}{cc} a & b \\ c & d \end{array}\right)\in Gl_2 (K)$, neither $a=d=0$ nor $b=c=0$ , and $E$ is a secret random elliptic curve from the isogeny class of a pairing friendly elliptic curve.  The trilinear map is publicized by specifying the following:
\begin{itemize}
\item
$\hat{\alpha}, \hat{\beta}\in \hat{E}[\ell]\subset K^{3n}$,
\item
the addition morphism $\hat{m}$ with the doubling map as a separate special case,
\item a set of maps of bounded locality: $\hat{\varphi}_i$, $i=1,\ldots, N$ where $N=O(n^2)$,
\item  two functions $\hat{g}$ and
$\hat{h}$ for the computation of $\hat{e}$.
\end{itemize}

\ \\The addition map and doubling map on $\hat{E}[\ell]$, as well as the maps  $\hat{\varphi}_i$, $i=1,\ldots, N$, will be be  specified using the methods developed in \S~\ref{ab}.
In \S~\ref{pairing} we will demonstrate
how $\hat{e}$ can be explicitly defined and specified, and efficiently computed using the specification of two functions $\hat{g}$ and
$\hat{h}$, both are products of semi-local functions of bounded degree.

\subsubsection{Security of blinding}
\ \\The following theorem addresses the security of blinding.  It is a simplified version of Theorem~\ref{specify-trilinear}, which will be proven later.
\begin{theorem}
\label{specify-trilinear1}
The information contained in the specification of the trilinear map can be described by $n^{O(1)}$ algebraic
conditions in $N$ unknowns where $N=n^{O(1)}$ and $N =\Omega(n^2)$. These algebraic conditions describe the relation between the published polynomials and the hidden polynomials, which
include the secret quadratic polynomials that determine the basic blinding map involved in the trilinear map construction.
Let $\tilde{V}_{{\cal T}}$ be the algebraic set determined by these conditions.
Let $\tilde{V}_{\la {\cal T} \ra}$ be the algebraic set determined by these conditions, however with the quadratic polynomials in the basic blinding map expressed in terms of the blinding parameters.
Then around every point of $\tilde{V}_{{\cal T}}$ (resp. $\tilde{V}_{\la {\cal T} \ra}$) an algebraic set of dimension $\Omega(n^2)$ (resp. $\Omega(n)$) can be embedded.
\end{theorem}

\ \\We remark that $\tilde{V}_{{\cal T}}$ and $\tilde{V}_{\la {\cal T} \ra}$ can be described in terms of generically constructed polynomial equations of bounded degrees with partial specializations determined by the coefficients of the published polynomials (see \S~\ref{specify-semilocal-sum}, \S~\ref{specify-semilocal-product}). Applying best known methods to find a point on $\tilde{V}_{{\cal T}}$ or $\tilde{V}_{\la {\cal T} \ra}$ , which by Theorem~\ref{specify-trilinear1} is  of positive dimension ($\Omega(n)$), takes expected time $2^{O(N\log N)}$ where $N$ is the number of unknown involved, and $N=\Omega(n^2)$ (\cite{Ay,HW}).   Therefore Theorem~\ref{specify-trilinear1}  provides strong evidence that the blinding is secure.  In addition, it will be shown in Theorem~\ref{specify-trilinear} that $\tilde{V}_{{\cal T}}$ and $\tilde{V}_{\la {\cal T} \ra}$ have additional properties which we call {\em triply confusing}.  They are also most likely {\em uniformly confusing} (see \S~\ref{semi-local-functions}).   These properties make it heuristically difficult to efficiently reduce sampling points on $\tilde{V}_{{\cal T}}$ or $\tilde{V}_{\la {\cal T} \ra}$ to sampling points on a zero dimensional set (see \S~\ref{semi-local-functions}), hence they provide additional evidence that efficient algorithms to find points on such algebraic sets seems unlikely to exist.

\subsubsection{Trapdoor discrete logarithm}
\label{trapdoor-dl}
\ \\Besides the security of blinding, the cryptographic strength of the trilinear map also depends on the hardness of the discrete logarithm problem on the third group $G_3$.  It is a trapdoor discrete logarithm problem which we describe below in more general terms since it may be of independent interest.

\ \\Let $E$ an elliptic curve defined over a finite field $K$.  We remark that for the problem defined here, $E$ is secret but not necessarily pairing friendly.
Let $\log |K|$, $\log\ell$ and $n$ are linear in the security parameter
The trapdoor secret consists of the following, described earlier in this section.
\begin{enumerate}
\item $\rho$ a randomly chosen blinding map,
\item $\hat{E}=\rho^{-1} \prod_{i=1}^n E_i$,
\item $M_1,\ldots,M_N \in Gl_n (\F_{\ell})$ with $N=O(n^2)$, such that each row of $M_i$ has two non-zero entries, which contain 1, and that the matrices $M_i$ together with the identity matrix $M_0$ generate $Mat_n (\F_{\ell})$ as a vector space over $\F_{\ell}$,
\item $R=\F_{\ell}[z_1,\ldots,z_N]$ a non-commutative associative algebra over free variables $z_1,\ldots,z_N$.
\end{enumerate}

\ \\The following are publicly specified:
\begin{enumerate}
\item $\hat{\beta}\in\hat{E}[\ell]$,
\item $\hat{\varphi}_1,\ldots,\hat{\varphi}_N\in\End (\hat{E}[\ell])$ (where $\varphi_i$ corresponds to $M_i$ as discussed before, both being secret),
\item $\hat{m}$ where $m:E\times E \to E$ is the secret addition morphism.
\end{enumerate}

\ \\The discrete logarithm problem is: Given $f\in R$ supported at 1, $z_1,\ldots,z_N$ and a monomial of degree $n$, to determine $a\in\F_{\ell}$ such that
$f(\hat{\varphi}_1,\ldots,\hat{\varphi}_N) (\hat{\beta}) = a \hat{\beta}$.  In other word, given $f\in [a]$ with $a$ unknown, the problem is to find $a$.

\ \\We assume that $n^{O(1)}$ many random samples of $[0]$ are publicly available.  The set $[0]$ contains at least $N^n$ linearly independent polynomials since there are $N^n$ (non-commutative) monomials of degree $n$.  Suppose $f\in [a]$. The probability that $f-a$ is linearly dependent on the $s=n^{O(1)}$ samples is negligibly small since $s << N^n$.   Therefore it seems very unlikely to mount an efficient linear algebra  attack, unless the trapdoor secret map $\lambda: R\to Mat_n (\F_{\ell})$ is revealed.

\ \\In the formulation above, it is not clear if the discrete logarithm problem can even be solved in subexponential time. In the setting of trilinear map, an efficiently computable pairing $\hat{e}$ between $G_1$ and $G_2$ is also made available.  Therefore the discrete logarithm problem can be solved in subexponential time after a reduction through $\hat{e}$ to the group $\mu_{\ell}\subset K$.

\subsubsection{The Decision-Diffie-Hellman (DDH) assumption}
\ \\It is easy to modify the trilinear map construction so that the Decision-Diffie-Hellman (DDH) assumption is conjecturally satisfied on the pairing groups.  We use a random blinding $\rho$ to construct $\hat{E}$, and form $\hat{\beta}$ and $G_2$ and $\hat{\varphi_i}$'s as above.  Then we use a different random blinding $\rho'$ to construct $\hat{E'} = {\rho'}^{-1} \prod_{i=1}^n E_i$, and form $\hat{\alpha}\in \hat{E'}[\ell]$ such that $\hat{\alpha}$ corresponds to $\alpha$ under
$\hat{E'}\stackrel{\rho'}{\to} \prod_{i=1}^n E_i \simeq E^n$. The pairing $\hat{e}$ is now between $\hat{E'}[\ell]$ and $\hat{E}[\ell]$.

\ \\We note that if the same blinding $\rho$ is used to form $G_1$ as before and $\hat{e}$ is a pairing between $\hat{E}[\ell]$ and $\hat{E}[\ell]$, the group $G_1$ may not satisfy the DDH assumption.  The reason is that we may heuristically assume $\hat{e} (\hat{\alpha}, \hat{\varphi}_i (\hat{\alpha})) \neq 1$ for some $i$, so we can use $\hat{\varphi}_i$ to induce a non-degenerate self-pairing on $G_1$.  We can verify $ab\hat{\alpha}$ from $a\hat{\alpha}$ and $b\hat{\alpha}$ using the following equality:
\[ \hat{e} (a\hat{\alpha},\hat{\varphi}_i ( b\hat{\alpha})) = \hat{e} (\hat{\alpha},\hat{\varphi}_i ( ab\hat{\alpha})).\]
When a different blinding $\rho'$ is used to form $\hat{E'}$, no map from $\hat{E'}[\ell]$ to $\hat{E}[\ell]$ is available to induce a self-pairing on $G_1$.  Similar remarks apply to $G_2$.

\section{Algebraic blinding}
\label{ab}
We continue our study of algebraic blinding techniques, especially techniques for specifying functions and maps.
As before we form a basic blinding map $\rho$ by choosing a random $\delta\in Gl_{3n} (K)$, and for $i=1,\ldots,n$, a random $\lambda_i=B_i\circ\lambda_{p_i,q_i}\circ A_i$  with $A_i,B_i\in Gl_3 (K)$ and  $p_i(x)\in K[x]$, $q_i(x,y)\in K[x,y]$ where $\deg p_i(x)=\deg q_i(x,y)=2$.

\ \\Write $\delta = (\delta_i)_{i=1}^n: \bar{k}^{3n}\to \prod_{i=1}^n \bar{k}^3$ with $\delta_i:\bar{k}^{3n}\to \bar{k}^3$ given by linear forms $L_{ij}$, $j=1,2,3$, in $3n$ variables.

\ \\Write   $\lambda_i (\bx) = (f_{ij} (\bx))_{j=1}^3$ where $\bx = (x,y,z)$.  The blinding space $W= \delta^{-1} \prod_{i=1}^n W_i$, where  $W_i$ is the zero set of $f_{i2}-f_{i3}$.

\ \\For $i=1,\ldots,n$,
$\lambda_i \circ \delta_i = (F_{ij})_{j=1}^3$ where  $F_{ij}=f_{ij}\circ (L_{i1}, L_{i2}, L_{i3})$ for $j=1,2,3$.
Let $\rho_i =pr\circ\lambda_i\circ\delta_i = (F_{i1}, F_{i2})$ where $pr$ denotes the projection $\bar{k}^3\to \bar{k}^2:(x,y,z)\to (x,y)$.
Then $W_i = \delta_i (W)$ and $W$ is the zero set of
$\{ F_{i2}-F_{i3} : i=1,\ldots,n\}$.

\ \\The basic blinding map associated with the blinding space $W$ is $\rho: \bar{k}^{3n} \to \prod_{i=1}^n \bar{k}^2$ where $\rho=(\rho_i)_{i=1}^n$.  We see that $\rho$ maps $W$ isomorphically to $\bar{k}^{2n}$.

\ \\{\bf Ideal of ambivalence} The ideal $I$ generated by $F_{i2}-F_{i3}$, $1\le i \le n$, is considered the ideal of ambivalence because for polynomials $H,H'$ such that $H-H'\in I$,   $H$ and $H'$ define the same map $W\to\bar{k}$.

\ \\Let $I_2$ be the submodule of $I$ generated by  $F_{i2}-F_{i3}$, $1\le i \le n$.

\begin{theorem}
\label{ambivalence}
Let $\rho=(\rho_i)_{i=1}^n:W\to \bar{k}^{2n}$ be a basic blinding map with $\rho_i = (F_{i1},F_{i2}) : W\to\bar{k}^2$ where $F_{ij}$ are quadratic polynomials in $3n$ variables for $i=1,\ldots,n$ and $j=1,2$.  Then $\rho_i = (H_{i1},H_{i2}): W\to\bar{k}^2$ for $i=1,\ldots n$ if $H_{ij}$ are polynomials such that $H_{ij} - F_{ij}\in I_2$ for  $i=1,\ldots,n$ and $j=1,2$.  The set of $(H_{ij})_{1\le i\le n;j=1,2}$  that determine the same basic blinding map as $\rho$ is isomorphic to ${\bar k}^{2 n^2}$.
\end{theorem}
\ \\{\bf Proof} The first assertion follows directly from the definition of $I$.
For the second assertion, since the zero set of $I$, which is $W$, has dimension $2n$, $F_{i2}-F_{i3}$, $i=1,\ldots,n$, are algebraically independent, hence linearly independent as well.  Hence there is a linear isomorphism between $I_2$ and $\bar{k}^{n}$.  From this the second and the third assertions follow. $\Box$%

\begin{lemma}
\label{Dalpha}
For $\alpha\in\bar{k}$, let $D_{\alpha} = \left(
\begin{array}{crr}
1 & \alpha & -\alpha\\
0& 1 & 0\\
0 & 0 &1
\end{array}
\right)$.  Consider a local quadratic isomorphism $\lambda=B\circ\lambda_{p,q}\circ A$ as described above.
Let $\lambda_{\alpha} = D_{\alpha}\circ\lambda=B_{\alpha}\circ\lambda_{p,q}\circ A$ where $B_{\alpha}=D_{\alpha} B$.
Then  $W_{\lambda} = W_{\lambda_{\alpha}}$ and $\lambda=\lambda_{\alpha}: W_{\lambda}\to \bar{k}^3$.
\end{lemma}
\ \\{\bf Proof}  Suppose $\lambda (\bx)= (f_1(\bx),f_2(\bx),f_3(\bx))$ for $\bx\in\bar{k}^3$ where $f_1,f_2,f_3$ are quadratic polynomials.  Then $\lambda_{\alpha}(\bx)= (f'_1(\bx),f_2(\bx),f_3(\bx))$ for $\bx\in\bar{k}^3$ where $f'_1 = f_1 + \alpha (f_2-f_3)$.  Therefore $W_{\lambda} = W_{\lambda_{\alpha}}$. $\Box$

\ \\Suppose we allow the blinding parameters to take values in $\bar{k}$, hence  $\delta\in Gl_{3n} (\bar{k})$, $A_i,B_i\in Gl_3 (\bar{k})$, $p_i$ and $q_i$ a quadratic polynomials with coefficients from $\bar{k}$ for $i=1,\ldots,n$.  Let $\la \rho \ra$ denote the set of parameters $(\delta, A_i,B_i,p_i,q_i:i=1,\ldots,n)$ that define the blinding map $\rho:W\to\bar{k}^{2n}$.

\begin{theorem}
\label{same-rho}
 Let $\rho:W\to \bar{k}^{2n}$ be a basic blinding map determined by parameters: $\delta\in Gl_{3n} (K)$, $\lambda_i = B_i\circ\lambda_{p_i,q_i}\circ A_i$, with $A_i,B_i\in Gl_3 (K)$, $p_i$ a quadratic polynomial in one variable, $q_i$ a quadratic polynomial in two variables, for $i=1,\ldots,n$.  Let $\lambda'_i = D_{\alpha_i}\circ\lambda_i$ with $\alpha_i\in\bar{k}$ for $i=1,\ldots,n$.  Let $\rho'=(\rho'_i)_{i=1}^n$ where $\rho'_i = pr\circ\lambda'_i\circ\delta_i$ for $i=1,\ldots,n$.  Then $\rho'=\rho:W\to\bar{k}^{2n}$, hence there is an injective map  $\bar{k}^{n}\to\la \rho\ra$.
\end{theorem}
\ \\{\bf Proof} The theorem follows immediately from Lemma~\ref{Dalpha}.  $\Box$.

\subsection{Semi-local functions}
\label{semi-local-functions}
We say that a rational function $f:\prod_{i=1}^n V_i \to \bar{k}$ defined over $K$ is $c$-{\em local} if there are $1\le i_1,\ldots,i_c \le n$ such that for $\bx=(x_i)$ with $x_i \in V_i$, $f(\bx)$ depends only on $x_{i_1},\ldots,x_{i_c}$.
We call $(i_1,\ldots,i_c)$ the {\em locality} of $f$.
We say that a rational function $f$ is of degree  $d$ if $f$ can be expressed as $\frac{G}{H}$ where $G$ and $H$ are polynomials and $d$ is the maximum of $\deg G$ and $\deg H$.
In this paper we only consider $c$-local functions of bounded but positive degree.  By abuse of notation we also write $f(\bx) = f(x_{i_1},\ldots,x_{i_c})$.

\ \\Suppose $f$ is $c$-local as above.  We consider the function
$g=f\circ\rho$ {\em semi-local}, noting that $g(\bx) = f(\rho_{i_1} (\bx), ...,\rho_{i_c} (\bx))$.
Denote by $[f]$ the set of $(h_1,h_2)$ where $h_1$ and $h_2$ are $2c$-variate polynomials such that
the degree of $h_1$ and $h_2$ is bounded by $d_j$ at $x_{i_j}$ for $j=1,\ldots,c$ and $f$ as a rational function on in $2c$ variables can be defined by $h_1/h_2$.  Denote by $[\rho]$ the set of $(H_{ij})_{i=1,\ldots,n; j=1,2}$ such that the basic blinding map $\rho$ can be defined by quadratic polynomials $H_{ij} (\bx)$, $i=1,\ldots,n$ and $j=1,2$, so that $\rho=(\rho_i)_{i=1}^n$ with $\rho_i = (H_{i1},H_{i2})$.  Then for $\bx\in W$, $g(\bx) = \frac{h_1 (H_{i_1,1} (\bx), H_{i_1, 2} (\bx),\ldots, H_{i_c,1} (\bx), H_{i_c,2}(\bx))}{h_2 (H_{i_1,1} (\bx), H_{i_1,2} (\bx),\ldots, H_{i_c,1} (\bx), H_{i_c,2}(\bx))}$.   We say that  $h_1,h_2,H_{11}, H_{12},\ldots, H_{n1},H_{n2}$ constitute a {\em semi-local decomposition} of $g$.  We denote such decomposition as $[f]\circ[\rho]$ where $[f]$ is the local part and $[\rho]$ the blinding part.

\ \\Let $A_i \in Gl_2 (\bar{k})$ for $i=1,\ldots,n$.  Let $A$ be the block-diagonal matrix with $A_1$, ..., $A_n$ as the diagonal blocks.  Let $g$ be a semi-local function as above.  If $g=f\circ\rho$ then
$g=(f\circ A^{-1})\circ (A \circ \rho)$.  Note that $f\circ A^{-1}$ has the same locality $c$ as $f$ at $i_1,\ldots,i_c$.
We say that $[f\circ A^{-1}]\circ [A \circ \rho]$ is obtained from $[f]\circ[\rho]$ by
the action of the matrix $A$.

\ \\We also consider rational functions $f: \prod_{i=1}^n V_i \times \prod_{i=1}^n V_i \to \bar{k}$ defined over $K$ that is local in the sense that for $\bx = (x_j)_{j=1}^n$ with $x_j\in V_j$ and $\by = (y_j)_{j=1}^n$ with $y_j\in V_j$, $f(\bx,\by)$ depends on $(x_i,y_i)$ for some $i$.  By abuse of notation we write $f(\bx,\by)=f(x_i,y_i)$. The function $g=f\circ(\rho,\rho):W\times W\to\bar{k}$ is {\em semi-local} in he sense that for $\bx,\by\in W$, $g(\bx,\by)=f(\rho_i (\bx), \rho_i (\by) )$.  Similar to the discussion before, a {\em semi-local decomposition} of $g$ is denoted
$[f]\circ[\rho]$ where the local part $[f]$
consists of $4$-variate polynomials $h_1$ and $h_2$, and the blinding part $\rho]$ consists of quadratic polynomials $H_{ij}$ in $3n$ variables,$i=1,\ldots,n$, $j=1,2$, such that for $\bx,\by\in W$, $g(\bx,\by) = \frac{h_1 (H_{i1} (\bx), H_{i2} (\bx), H_{i1} (\by), H_{i2}(\by))}{h_2 (H_{i1} (\bx), H_{i2} (\bx), H_{i1} (\by), H_{i2}(\by))}$.
Let $A_j \in Gl_2 (\bar{k})$ for $j=1,\ldots,n$.  Let $A$ be the block-diagonal matrix with $A_1$, ..., $A_n$ as the diagonal blocks.  Then  we similarly obtain semi-local decomposition $[f\circ A^{-1}]\circ [A \circ \rho]$ from $[f]\circ [\rho]$ by the action of $A$.

\ \\Similar consideration can be made if $f$ maps $V_i\times V_i\times V_i \to \bar{k}$, or more generally if $f:V_i^c \to \bar{k}$ where $c$ is a constant.

\begin{theorem}
\label{semi-local-thm}
Suppose we have a set of semi-local functions $g_i$, $i=1,\ldots,m$, such that $g_i$ has semi-local decomposition
$[f_i]\circ [\rho]$ for all $i$, where $f_i$ is a local function and $\rho$ is a basic blinding map.
Then the following hold.
\begin{enumerate}
\item
There is an injective map $\bar{k}^{2n^2}\to [\rho]$.  More explicitly if $(F_{ij})_{i=1,\ldots,n; j=1,2}$ define $\rho$, then so does $(F'_{ij})_{i=1,\ldots,n; j=1,2}$ if $F'_{ij}-F_{ij}\in I_2$.
\item There is an injective map  $\bar{k}^{n}\to\la \rho\ra$.
\item There is an injective map $\bar{k}\to [f_i]$ if $f_i$ is $c$-local depending on $V_{i_1}\times\ldots V_{i_c}$ and for some $j$ the degree of $f_i$ at $x_{i_j}$ is greater to equal to the minimum degree of polynomials in the ideal defining $V_{i_j}$.
\item
Let $A_j \in Gl_2 (\bar{k})$ for $j=1,\ldots,n$.  Let $A$ be the block-diagonal matrix with $A_1$, ..., $A_n$ as the diagonal blocks.  Then $g_i$ has semi-local decomposition $[f_i\circ A^{-1}]\circ [A \circ \rho]$ for $i=1,\ldots,m$.
\end{enumerate}
\end{theorem}
\ \\{\bf Proof} The first assertion follows from Theorem~\ref{ambivalence}.  The second assertion follows from Theorem~\ref{same-rho}. For the third assertion observe that if $(h_1,h_2)\in [f_i]$ then $(h'_1,h'_2)\in [f_i]$ if $h'_1 - h_1$ and $h'_2-h_2$ are in the ideal defining $V_{i_j}$.  The last assertion follows the discussion above.  $\Box$

\ \\Let $S=\{g_i: i=1,\ldots,m\}$ be as in the theorem.  Let $V_S$ be the union of $[f_1]\times\ldots\times [f_m]\times [\rho]$, where the union is over all $f_1,\ldots,f_m,\rho$ such that $g_i$ has semi-local decomposition $[f_i]\circ [\rho]$, $i=1,\ldots,m$.  Similarly let $V_{\la S\ra}$ be the union of $[f_1]\times\ldots\times [f_m]\times \la\rho\ra$, where the union is over all $f_1,\ldots,f_m,\rho$ such that $g_i$ has semi-local decomposition $[f_i]\circ [ \rho ]$, $i=1,\ldots,m$.

\ \\We say that $V_S$ and $V_{\la S\ra}$ are {\em triply confusing} for the following reasons:
(1) the first two assertions of the theorem states that $V_S$ (resp. $V_{\la S \ra}$) is confusing in the blinding part $[\rho]$ (resp. $\la \rho \ra$),
(2) the third assertion states that  $V_S$ (resp. $V_{\la S\ra}$) is confusing in the local part, and
(3) the fourth assertion states that there is $Gl$-action on the decomposition $[f_i]\circ [\rho]$.

\ \\To understand the utility of the triply confusing property, consider the simple case $S$ consists of a semi-local function $g$ with semi-local decomposition $[f]\circ [\rho]$, $g=f\circ (F_1, F_2)$ on $W$, where $f$ is a bivariate quadratic polynomial, and $F_1$, $F_2$  are quadratic in $3n$ variables.  If we ignore the other parts of the basic blinding map $\rho$ but focus on $F_1$ and $F_2$,  then $V_S$ simplifies to an algebraic set defined in terms of the $N=O(n^2)$ unknown coefficients of $f$ and $F_1$ and $F_2$, and $V_S$ is uniformly of positive dimension $\Omega(n)$.  Consider the problem of sampling points on $V_S$ assuming $V_S$ is known. Sampling points on $V_S$ can be done by an exponential time  reduction to sampling points on a hypersurface (of exponential degree) (see \cite{HW}).  Should an efficient sampling of points on $V_S$ exists, it would likely involve an efficient reduction to sampling points on some efficient-to-sample zero dimensional set.  Is this possible?  The following heuristic analysis suggests this is unlikely.

\ \\Note that that $g=f \circ (F_1, F_2)$ not as global functions, but as functions on the blinding space $W$.  In this case we may write $g\equiv f \circ (F_1, F_2)\mod I$ where $I$ is the ideal of ambivalence.  If we consider the equality as being global for the moment, and assume $g$ is known, then the unknown coefficients of $f$ and $F_1, F_2$ satisfies $n^{O(1)}$ very sparse conditions, resulting in a sparse polynomial system of bounded degree.  However this is not the case with our situation.  In our case the $\mod I$-condition leads to a much denser formulation of $V_S$ where each polynomial equation contains $\Omega(n)$ non-trivial terms.
To reduce dimension in the current situation one can try to pick a unique member from the class $[F_1]$ heuristically by imposing linear conditions on the coefficients of $F_1$, and similarly for $F_2$ and $f$.  Imposing such conditions reduces the dimension (while making the polynomial system even more dense).  However after imposing such conditions the polynomial system remains in positive dimension due to the $Gl_2$-action; that is, if $g$ has decomposition $[f]\circ [(F_1, F_2)]$ then $g$ also has decomposition $[f\circ A]\circ [ A^{-1} (F_1, F_2)]$ for $A\in Gl_2 (\bar{k})$.  Therefore it seems unlikely that sampling points on $V_S$ can be efficiently reduced to a zero-dimensional case given the triply confusing property.

\ \\We shall see later on when we specify functions and maps on $\hat{V}$,
the relationship between the published entities and the hidden entities is described by certain algebraic sets.
When a set $S$ os semi-local functions is involved in such a specification, we shall see that $V_S$ and $V_{\la S\ra}$  are locally embedded in the algebraic sets, making them triply confusing as well.

\ \\It is desirable for $V_S$ to be not only triply confusing, but uniformly confusing in the sense that there is not a special member in the $Gl_2$-orbit of a local function $f_i$.  Consider for example $V_i$ is the image of some secret elliptic curve $E$ of the form $y^2=x^3+ax+b$, under random choices of $A\in Gl_2 (K)$, and $g_i$ is determined by the doubling morphism.  Then $f_i$ is typically a dense bivariate rational function.  However there is some $A\in Gl_2 (K)$ such that $f_i\circ A$ is the doubling morphism on $E$, which takes a much simpler form.

\ \\To make $V_S$ uniformly and triply confusing in the above situation, we apply the following treatment.  Let $\jmath$ be the map on $\bar{k}^{*2}$ such that $\jmath (x,y)=(x^{-1}, y^{-1})$.  For an algebraic set $V$, we let $\overline{V}=\jmath (V)$, $V^A = A(V)$ for $A\in Gl_2 (K)$, hence $\overline{V^A} = \jmath{A(V)}$.  For an algebraic set $V$ that is the base algebraic set to be blinded,  we consider the set $\{ \overline{V^A} : A=\left( \begin{array}{cc} a & b \\ c & d \end{array}\right)\in Gl_2 (K)$, neither  $a=d=0$ nor $bc\neq 0 \}$.  We choose  $V_i$ uniformly and randomly from the set.

\ \\Suppose $f_0: V \to \bar{k}$, a local function. If $f_0$ is known to be of a special form then $\overline{f}_0 = f_0 \circ\jmath$ is also of a special form. For $A\in Gl_2 (K)$, the corresponding function of $f_0$  on $\overline{V^A}$ is $\overline{f}_A=f_0\circ A^{-1} \circ \jmath$.  Now suppose $f_0=g/h$ where $g$ and $h$ are bivariate polynomials, then for random $A$ with all nonzero entries,
$\overline{f}_A$ is of the form $(xy)^{\deg h - \deg g} g' / h'$ where $g'$ (resp. $h'$) is a dense polynomials of the same degree as $g$ (resp. $h$).   Below we argue that the orbits of $\overline{f}_A$ under $Gl_2 (K)$ are likely all disjoint with different choices of $A$, and in particular it is unlikely for $\overline{f}_0$ to be in the orbit of $\overline{f}_A$. Otherwise $\overline{f}_A\circ B = \overline{f}_C$ for some $B,C\in Gl_2 (K)$, hence
$f_0\circ A^{-1}\circ \jmath\circ B = f_0\circ C^{-1}\circ \jmath$, which is unlikely unless
$ A^{-1}\circ \jmath\circ B = C^{-1}\circ \jmath$, hence
$ B  = \jmath AC^{-1} \jmath$, but this would contradict the following lemma.

\begin{lemma}
\label{j}
For $A=\left( \begin{array}{cc} a & b \\ c & d \end{array}\right)\in Gl_2 (K)$, $\jmath \circ A \circ \jmath \in Gl_2 (K)$ only if $b=c=0$ or $a=d=0$.
\end{lemma}
\ \\{\bf Proof} Let $A=\left( \begin{array}{cc} a & b \\ c & d \end{array}\right)$ and
$B=\left( \begin{array}{cc} r & s \\ t & u \end{array}\right)$.  Then $\jmath \circ A \circ \jmath =B$ implies for all $x,y\in \bar{k}^*$, $(ax+by)^{-1} = r x^{-1}+s y^{-1}$ and $(cx+dy)^{-1} = tx^{-1} + u y^{-1}$.  It follows by simple algebra that $as=br=0$.   Likewise $cu=dt=0$, and the assertion follows. $\Box$

\ \\Consider now a semi-local function $g$ with semi-local decomposition $f\circ\rho$ where $f=\overline{f}_A=f_0\circ A^{-1} \circ \jmath$.  We have argued that $V_g$ is not only triply confusing but most likely uniformly confusing as well. In this kind of situation where $\jmath$ is involved and some information is known to the public about $f_0$ we may also consider a more refined semi-local decomposition of $g$ of the form $f' \circ \jmath\circ\rho$ where $f'= f_0\circ A^{-1}$.  In this case the triply confusing property stated in Theorem~\ref{semi-local-thm} still holds except the $Gl$-action needs to be modified as the action by a subgroup.  More explicitly we have the following.
\begin{enumerate}
\item
There is an injective map $\bar{k}^{2n^2}\to [\rho]$.
\item There is an injective map  $\bar{k}^{n}\to\la \rho\ra$.
\item There is an injective map $\bar{k}\to [f']$ if $f'$ is $c$-local depending on $V_{i_1}\times\ldots V_{i_c}$ and for some $j$ the degree of $f'$ at $x_{i_j}$ is greater to equal to the minimum degree of polynomials in the ideal defining $V_{i_j}$.
\item
Let $A_j \in Gl_2 (\bar{k})$ be either of the form $\left( \begin{array}{cc} a & 0 \\ 0 & b \end{array}\right)$ or of the form $\left( \begin{array}{cc} 0 & a \\ b & 0 \end{array}\right)$, for $j=1,\ldots,n$.  Let $A$ be the block-diagonal matrix with $A_1$, ..., $A_n$ as the diagonal blocks.  Then $g$ has semi-local decomposition $[f'\circ A']\circ [A \circ \rho]$ for $i=1,\ldots,m$.  Here for $D=\left( \begin{array}{cc} a & 0 \\ 0 & b \end{array}\right)$, $D'=\left( \begin{array}{cc} a^{-1} & 0 \\ 0 & b^{-1} \end{array}\right)$; for  $D=\left( \begin{array}{cc} 0 & a \\ b & 0 \end{array}\right)$, $D'=\left( \begin{array}{cc} 0 & a^{-1}  \\  b^{-1} & 0 \end{array}\right)$
\end{enumerate}

\ \\The last property follows from the identity that $A'\circ\jmath = \jmath\circ A$ if $A$ is either of the form $\left( \begin{array}{cc} a & 0 \\ 0 & b \end{array}\right)$ or of the form $\left( \begin{array}{cc} 0 & a \\ b & 0 \end{array}\right)$.

\subsection{Specifying a semi-local sum}
\label{specify-semilocal-sum}
\ \\For $i > 0$, let $I_i$ be the submodule of $I$ generated by elements of the form $t (F_{i2}-F_{i3})$ where $t$ is a monomial of degree at most $i-2$.  If a map $W\to\bar{k}$ can de defined a polynomial $h$ of degree $d$, then it is also defined by any polynomial in $h+I_d$.
For a random choice of basic blinding map $\rho$, the associated $F_{ij}$ are dense quadratic polynomials in $\bx$, so are $F_{i2}-F_{i3}$.  Therefore a random element of $h+I_d$ is likely a dense polynomial of degree $d$ in $\bx$.
 We write $f\in_R h+I_d$ to denote a uniform random selection from $h+I_d$.

\ \\To specify a function $f$ that is the sum of hidden semi-local functions, say $f=\sum_{i=1}^m \varphi_i$ where $\varphi_i$ is a hidden semi-local function, we take the following steps to specify $f$ as a sum of $m$ random-looking functions that are not semi-local.  Again we focus on the case $\varphi_i = f_i\circ\rho$ where $f_i$ is $c$-local in that it depends on $V_{i_1}\times\ldots\times V_{i_c}$ for some $i_1,\ldots,i_c$.  The case that $\varphi_i = f_i \circ (\rho,\rho)$ or $\varphi_i = f_i \circ (\rho,\rho,\rho)$ where $f_i$ is local depending on $V_i\times V_i$ or $V_i\times V_i\times V_i$ is similar.
\begin{enumerate}
\item Construct $2m$ random linear forms $\ell_{i,j} (\bx)$, $i=1,\ldots,m$, $j=1,2$.  Put $\ell_{m+1,j} = \ell_1,j$ for $j=1,2$.
\item Suppose $\varphi_i (\bx) = \frac{g_i (\bx)}{h_i (\bx)}$ on $W$ with $g_i,h_i \in K[\bx]$.   Then
$\frac{g_i}{h_i}+ \frac{\ell_{i,1}}{\ell_{i,2}} - \frac{\ell_{i+1,1}}{\ell_{i+1,2}} = \frac{g'_i}{h'_i}$ where
$h'_i = h_i \ell_{i,2} \ell_{i+1,2}$ and $g'_i = g_i \ell_{i,2}\ell_{i+1,2} +\ell_{i,1} h_i \ell_{i+1,2} - \ell_{i+1,1}h_i\ell_{i,2}$.
\item  Choose random $g''_i\in_R g'_i +I_{d_{i,1}+2}$ where $d_{i,1}=\deg g'_i$ and $h''_i\in_R h'_i +I_{d_{i,2}+2}$ where
$d_{i,2} =\deg h'_i$ for $i=1,\ldots,m$.
\item Publish $\{ g''_i, h''_i : i=1,\ldots,m\}$, and specify $\varphi$ as $\sum_{i=1}^m \frac{g''_i}{h''_i}$  on $W$.  (Note that $\sum_{i=1}^m \frac{\ell_{i,i}}{\ell_{i,2}} -  \frac{\ell_{i+1,i}}{\ell_{i+1,2}}= 0$.)
\end{enumerate}

\ \\Write $f=_{W} g$ if for rational functions $f$ and $g$ on $\bar{k}^{3n}$, $f(\bx)=g(\bx)$ for all $\bx\in W$.

\ \\For simplicity assume $f_i$ has locality 2 and depends on $V_{i_1}\times V_{i_2}$. Let
${\bf F}_i = (F_{i_1,1},F_{i_1,2},F_{i_2,1},F_{i_2,2})$.  Suppose $f_i = \frac{h_{i1}}{h_{i2}}$ where $h_{i1}$ and $h_{i2}$ are polynomials in 4 variables.

\ \\
For $i=1,\ldots,m$,
\begin{eqnarray}
\label{lij-sum}
f_i \circ (\rho_{i_1}, \rho_{i_2}) +\frac{\ell_{i,1}}{\ell_{i,2}} - \frac{\ell_{i+1,1}}{\ell_{i+1,2}} \\
=\frac{h_{i1} \circ {\bf F}_i}{h_{i2} \circ {\bf F}_i} +\frac{\ell_{i,1}}{\ell_{i,2}} - \frac{\ell_{i+1,1}}{\ell_{i+1,2}} =_W \frac{g''_i}{h''_i}.
\end{eqnarray}
\ \\Let $U,V$ be two algebraic sets.  We say that $U$ is locally embedded around a point $\alpha$ of $V$ if there is an injective morphism $\iota: U\to V$ such that $\alpha\in \iota (U)$.

\ \\The equation~(\ref{lij-sum}) characterizes the algebraic condition determined by $g''_i$ and $h''_i$ in relation to the unknown $h_{i1}$, $h_{i2}$,  ${\bf F}_i$, and $\ell_{i,j}$ and $\ell_{i+1,j}$, $j=1,2$.  Note that if we treat the coefficients of $g''_i$ and $h''_i$ also as unknown for the moment, then the algebraic condition can be expressed generically by polynomial equations in $x_1,\ldots,x_{3n}$ and all the unknown coefficients involved, followed by specialization at the coefficients of $g''_i$ and $h''_i$.
The equation~(\ref{lij-sum}) describes an algebraic set $V$ where each point $\alpha$ contains the $O(n^2)$ coefficients of some $f_i$, ${\bf F}_i$, $\ell_{i,j}$ and $\ell_{i+1,j}$, $j=1,2$, that satisfy the equation.
The subset of coordinates of $\alpha$ describing $f_i$ and ${\bf F}_i$ corresponds to a point in $V_g$ where $g = f_i\circ (\rho_{i_1},\rho_{i_2})$, a semi-local function.  The set $V_g$ can be embedded around $\alpha$.  So $V$
has dimension $\Omega (n)$ where around every point a triply confusing algebraic set of dimension $\Omega(n)$, which is likely uniformly confusing as well, can be embedded.

\ \\The information contained in the specification is described by $m$ algebraic conditions of the form~(\ref{lij-sum}) giving relations of the specifying polynomials $g''_i$, $h''_i$, $i=1,\ldots,m$, to the hidden polynomials including $h_{i1}$, $h_{i2}$, $F_{ij}$, and $\ell_{ij}$.

\ \\Let $V_f$ be the algebraic set determined by these $m$ conditions. Let $V_{\la f\ra}$ be the algebraic set determined also by these conditions, however with $F_{ij}$ expressed in terms of the blinding parameters.

\ \\Let $U,V$ be two algebraic sets.  We say that $U$ is locally embedded around point $\alpha$ of $V$ if there is an injective morphism $\iota: U\to V$ such that $\alpha\in \iota (U)$.

\ \\A point $\alpha$ of $V_f$ (resp. $V_{\la f\ra}$) determines some $h_{i1}$, $h_{i2}$, $F_{ij}$, and $\ell_{ij}$ that satisfy the $m$ algebraic conditions of the form~(\ref{lij-sum}).  The
local functions $f_i=\frac{h_{i1}}{h_{i2}}$ has the same locality and local degree as the local function involved in $\varphi_i$, $i=1,\ldots,m$.   The $F_{ij}$ determine a basic blinding maps $\rho=(\rho_i)_{i=1}^n$ where
$\rho_i = (F_{i_1},F_{i2})$.  Let $\varphi'_i$ be the semi-local function with semi-local decomposition $[f_i]\circ [\rho]$ for $i=1,\ldots,m$.  Suppose $\varphi'_i$ also has semi-local decomposition $[f'_i]\circ [\rho']$.  Then every  $(h'_{i1}, h'_{i2})\in [f'_i]$, $i=1,\ldots,m$, and $(F'_{ij})\in [\rho']$ also satisfy the equations with $\ell_{i,j}$ and $\ell_{i+1,j}$.  Let $S'=\{\varphi'_i:i=1,\ldots,m\}$.
We have injective maps
\begin{eqnarray*}
V_{S'}=\cup [f_1]\times\ldots\times [f_m]\times [\rho] \to \cup [f_1]\times\ldots\times [f_m]\times [\rho]\times \{(\ell_{ij})\}\subset V_f\\
V_{\la S'\ra}=
\cup [f_1]\times\ldots\times [f_m]\times \la\rho\ra \to \cup [f_1]\times\ldots\times [f_m]\times \la\rho\ra\times \{(\ell_{ij})\}\subset V_{\la f\ra}.
\end{eqnarray*}
It follows that $V_{S'}$ (resp. $V_{\la S'\ra}$) is embed around $\alpha$.

\ \\We have proven the following:

\begin{theorem}
\label{semi-local-sum-thm}
Let $f:W\to\bar{k}$ be a function which is the sum of $m$ semi-local functions of bounded degree.  Suppose $f$ is specified by a set of $2m$ polynomials $g''_i$, $h''_i$, $i=1,\ldots,m$ by following the procedure described above.
The information contained in the specification can be described by a set of $m$ algebraic conditions of the form~(\ref{lij-sum}) giving relations of the published polynomials $g''_i$, $h''_i$, $i=1,\ldots,m$, to the hidden polynomials including $h_{i1}$, $h_{i2}$, $F_{ij}$, and $\ell_{ij}$.
Let $V_f$ be the algebraic set determined by these $m$ conditions. Let $V_{\la f\ra}$ be the algebraic set determined also by these conditions, however with $F_{ij}$ expressed in terms of the blinding parameters.  Every point of $V_f$ (resp. $V_{\la f\ra}$) determines, through a subset of its coordinates, a set of semi-local functions of the same type as the $m$ semi-local functions that sum to $f$.  Suppose $\alpha$ is a point in $V_f$ (resp. $V_{\la f\ra}$) and $S'$ is the set of semi-local functions determined by $\alpha$.
Then $V_{S'}$ (resp. $V_{\la S'\ra}$) is locally embedded around $\alpha$.
\end{theorem}

\subsection{Specifying a semi-local product}
\label{specify-semilocal-product}
\ \\Similarly, to specify a function $f$ that is the product of semi-local functions, say $f=\prod_{i=1}^m \varphi_i$ where $\varphi_i$ is semi-local, we take the following steps to specify $f$ as a product of $m$ random-looking functions that are not semi-local. Again we focus on the case $\varphi_i = f_i\circ\rho$ where $f_i$ is $c$-local in that it non-trivially depends on $V_{i_1}\times\ldots\times V_{i_c}$ for some $i_1,\ldots,i_c$.  The case that $\varphi_i = f_i \circ (\rho,\rho)$ or $\varphi_i = f_i \circ (\rho,\rho,\rho)$ where $f_i$ is local depending on $V_i\times V_i$ or $V_i\times V_i\times V_i$ is similar.
\begin{enumerate}
\item Construct $m$ random linear forms $\ell_{i} (\bx)$, $i=1,\ldots,m$, $j=1,2$.  Put $\ell_{m+1} = \ell_1$.
\item Suppose $\varphi_i (\bx) = \frac{g_i (\bx)}{h_i (\bx)}$ on $W$ with $g_i,h_i \in K[\bx]$.   Then
$\frac{g_i}{h_i} \frac{\ell_{i}}{\ell_{i+1}}= \frac{g'_i}{h'_i}$ where
$h'_i = h_i \ell_{i+1}$ and $g'_i = g_i \ell_{i}$.
\item  Choose random $g''_i\in_R g'_i +I_{d_{i,1}+1}$ where $d_{i,1}=\deg g'_i$ and $h''_i\in_R h'_i +I_{d_{i,2}+1}$ where
$d_{i,2} =\deg h'_i$ for $i=1,\ldots,m$.
\item Publish $\{ g''_i, h''_i : i=1,\ldots,m\}$, and specify $\varphi$ as $\prod_{i=1}^m \frac{g''_i}{h''_i}$  on $W$.  (Note that $\prod_{i=1}^m \frac{\ell_{i}}{\ell_{i+1}} = 1$.)
\end{enumerate}

\ \\Suppose $f_i = \frac{h_{i1}}{h_{i2}}$ where $h_{i1}$ and $h_{i2}$ are polynomials in 4 variables.  Similar to the case in \S~\ref{specify-semilocal-sum}, the information contained in the specification of $f$ is captured in $m$ algebraic conditions, $i=1,\ldots,m$:
\begin{eqnarray}
\label{lij-product}
(\frac{h_{i1} \circ {\bf F}_i}{h_{i2} \circ {\bf F}_i})  \frac{\ell_{i}}{\ell_{i+1}}  =_W \frac{g''_i}{h''_i}.
\end{eqnarray}
As before, if we treat the coefficients of $g''_i$ and $h''_i$ also as unknown for the moment, then the algebraic condition can be expressed generically by polynomial equations in $x_1,\ldots,x_{3n}$ and all the unknown coefficients involved, followed by specialization at the coefficients of $g''_i$ and $h''_i$.
\ \\The following theorem can be proven in a way that is similar to the proof for Theorem~\ref{semi-local-sum-thm}.
\begin{theorem}
\label{semi-local-product-thm}
Let $f:W\to\bar{k}$ be a function which is the product of $m$ semi-local functions of bounded degree.  Suppose $f$ is specified by a set of $2m$ polynomials $g''_i$, $h''_i$, $i=1,\ldots,m$ by following the procedure described above.
The information contained in the specification can be described by a set of $m$ algebraic conditions of the form~(\ref{lij-product}) giving relations of the specifying polynomials $g''_i$, $h''_i$, $i=1,\ldots,m$, to the hidden polynomials including $h_{i1}$, $h_{i2}$, $F_{ij}$, and $\ell_{i}$.
Let $V_f$ be the algebraic set determined by these $m$ conditions. Let $V_{\la f\ra}$ be the algebraic set determined also by these conditions, however with $F_{ij}$ expressed in terms of the blinding parameters.
Every point of $V_f$ (resp. $V_{\la f\ra}$) determines, through a subset of its coordinates, a set of semi-local functions of the same type as the $m$ semi-local functions that multiply to $f$.  Suppose $\alpha$ is a point in $V_f$ (resp. $V_{\la f\ra}$) and $S'$ is the set of semi-local functions determined by $\alpha$.
Then $V_{S'}$ (resp. $V_{\la S'\ra}$) is locally embedded around $\alpha$.
\end{theorem}

\subsection{Specifying maps of bounded locality}
\ \\We consider two kinds of rational maps of bounded locality that will be involved in the trilinear map construction.

\ \\First suppose $\varphi=(\varphi_i)_{i=1}^n:\prod_{i=1}^n V_i \to \prod_{i=1}^n V_i$ is a rational map of locality bounded by $c$ in the sense that for all $i$, $\varphi_i: \prod_{j=1}^n V_j \to  V_i\subset\bar{k}^2$ consists of two $c_i$-local functions with $c_i \le c$. For simplicity of discussion we assume $\varphi$ has locality 2.  Applying the basic blinding map $\rho$ to $\varphi$ we get $\hat{\varphi}: W \to W$ such that $\rho\hat{\varphi}=\varphi\circ\rho$.

\ \\Second suppose $\varphi=(\varphi_i)_{i=1}^n:\prod_{i=1}^n V_i \times \prod_{i=1}^n V_i\to \prod_{i=1}^n V_i$ is a rational map such that for all $i$, $\varphi_i: \prod_{i=j}^n V_j \times \prod_{j=1}^n V_j\to V_i$ is local in that it depends only on $V_i\times V_i$.  Applying the basic blinding map $\rho$ to $\varphi$ we get $\hat{\varphi}:W\times W\to W$ such that $\rho\hat{\varphi}=\varphi\circ(\rho,\rho)$.

\ \\Let $\hat{\varphi} = (\hat{\varphi}_i)_{i=1}^{3n}$.  In the proofs of the results in this section we will focus on the first case  where $\varphi=(\varphi_i)_{i=1}^n:\prod_{i=1}^n V_i \to \prod_{i=1}^n V_i$.  The argument for the second case is very similar and is omitted.

\ \\We adopt the following notation.  Suppose $f_i$, $i=1,2,m$, are polynomials in 3 variables, and $g_i$, $i=1,2,3$, are polynomials.  Let ${\bf f} = (f_1,\ldots,,f_m)$ and ${\bf g} = (g_1,g_2,g_3)$.  Then
\[
{\bf f} \circ {\bf g} := (f_1,\ldots, f_m)\circ (g_1,g_2,g_3) := (f_1 (g_1,g_2,g_3), \ldots, f_m (g_1,g_2,g_3)).
\]

\ \\As before we have
\begin{eqnarray*}
\delta_i = (L_{i1},L_{i2},L_{i3})\\
\lambda_i = (f_{i1},f_{i2},f_{i3})\\
F_{ij} = f_{ij}\circ \delta_i\\
\rho_i = pr\circ\lambda_i\delta_i = (F_{i1},F_{i2})
\end{eqnarray*}

\ \\Suppose $\varphi_i$ determined by $V_{i_1}\times V_{i_2}$.  Let $g_{ij},g'_{ij}$ be polynomials in 4 variables such that $\varphi_i (\bx) = \frac{g_{ij}(x_{i_1},x_{i_2})}{g'_{ij}(x_{i_1},x_{i_2})}$ where $\bx=(x_j)_{j=1}^n\in\prod_{j=1}^n V_j$ with $x_j\in V_j \subset\bar{k}^2$.

\ \\Let ${\bf F}_i = (F_{i_1,1},F_{i_1,2}, F_{i_2,1},F_{i_2,2})$.  Then
\[ \varphi_i \circ\rho = (\frac{g_{i1}}{g'_{i1}},\frac{g_{i2}}{g'_{i2}} )\circ {\bf F}_i .\]

\ \\Let $\mu_i = \lambda_i^{-1} = (h_{i1},h_{i2},h_{i3})$.  Let $\tilde{\mu}_i (x,y) = \mu_i (x,y,y)$.
Let
\[{\bf u}_i=(u_{i1},u_{i2},u_{i3})= \tilde{\mu}_i\circ\varphi_i=(h_{i1},h_{i2},h_{i3})\circ (\frac{g_{i1}}{g'_{i1}},\frac{g_{i2}}{g'_{i2}},\frac{g_{i2}}{g'_{i2}} ).\]
\ \\Then $\tilde{\mu}_i\circ\varphi_i\circ\rho = {\bf u}_i \circ {\bf F}_i$
is semi-local.
Let $(v_i)_{i=1}^{3n}$ be such that $u_{ij}\circ {\bf F}_i = v_{3(i-1)+j}$.
Then $\hat{\varphi}_i = \delta^{-1} (v_i)_{i=1}^{3n}$.  We have proved the following
\begin{lemma}
For $i=1,\ldots, 3n$, $\hat{\varphi}_i$ is the sum of $3n$ semi-local functions.
\end{lemma}

\ \\We apply the procedure described in \S~\ref{specify-semilocal-sum} to specify each $\hat{\varphi}_i$  as the sum of $3n$ random looking functions from $W$ to $\bar{k}$.  Suppose $\hat{\varphi}_i$ is specified as $\sum_{j=1}^{3n} \frac{G_{ij}}{G'_{ij}}$ for some polynomials $G_{ij}$ and $G'_{ij}$.   Therefore $\hat{\varphi}$ is specified by these
$O(n^2)$ polynomials $G_{ij}$ and $G'_{ij}$.

\ \\Again, write $f=_{W} g$ if for rational functions $f$ and $g$ on $\bar{k}^{3n}$, $f(\bx)=g(\bx)$ for all $\bx\in W$.
Now consider information revealed by the specification of $\hat{\varphi}$ about the hidden map $\varphi$ and the blinding parameters $\delta$, $\lambda_i$, $i=1,\ldots,n$.

\ \\Suppose $\delta^{-1} = (w_{ij})_{1\le i, j \le 3n}$.  We have
$\hat{\varphi}_i=\sum_{j=1}^{3n} w_{ij} v_j$ where $v_j =u_{rs} \circ {\bf F}_r$ with
$1\le r\le n$ and $1\le s \le 3$ such that $j=3(r-1)+s$.

\ \\We have
$w_{ij}v_j + \frac{\ell_{i,j}}{\ell'_{i,j}} -  \frac{\ell_{i,j+1}}{\ell'_{i,j+1}}=_W \frac{G_{ij}}{G'_{ij}}$, and $\ell_{ij}$, $\ell'_{ij}$ are linear forms for $i,j=1,\ldots,3n$ with $\ell_{i,3n+1}=\ell_{i,1}$ and $\ell'_{i,3n+1}=\ell'_{i,1}$.

\ \\Put $w_{ij} u_{r_j,s_j} = \frac{g}{g'}$ where $g,g'$ are polynomials in 4 variables.  Then as  Theorem~\ref{semi-local-sum-thm} is applied in this situation we get
\begin{eqnarray}
\label{Gij}
\frac{g\circ {\bf F}_{r_j}}{g'\circ {\bf F}_{r_j}}+\frac{\ell_{ij}}{\ell'_{ij}} -  \frac{\ell_{i,j+1}}{\ell'_{i,j+1}}=_W \frac{G_{ij}}{G'_{ij}}
\end{eqnarray}
Equation~(\ref{Gij}) characterizes the condition in specifying $G_{ij}$ and $G'_{ij}$, where $\ell_{ij}$ are unknown linear forms in $3n$ variables, $F_{ij}$ are unknown quadratic polynomials in $3n$ variables, $g$ and $g'$ are unknown polynomials in 4 variables of degree $2\deg\varphi_i = O(1)$, where $w_{ij} u_{r_j,s_j} = \frac{g}{g'}$.

\ \\The other kind of information provided once $\hat{\varphi}$ is specified is the information that may be obtained by evaluation of $\hat{\varphi}$ on $W$.    Suppose $\alpha\in W$ and $\hat{\varphi} (\alpha) = \beta$.
Then for $i=1,\ldots, n$, we have $\varphi_i\circ\rho (\alpha)=\rho (\beta)$, hence we get the condition
\begin{eqnarray}
\label{gFF1}
 \frac{g_{i1} ({\bf F}_i (\alpha))}{g'_{i1} ({\bf F}_i (\alpha))} = F_{i1}(\beta)\\
 \label{gFF2}
\frac{g_{i2} ({\bf F}_i (\alpha))}{g'_{i2} ({\bf F}_i (\alpha))} = F_{i2}(\beta)
\end{eqnarray}

\ \\The following two theorems can be proved in a way that is similar to the proof or Theorem~\ref{semi-local-sum-thm}.

\begin{theorem}
\label{specify-hat}
Suppose $\hat{\varphi}$ is specified using the procedure described in \S~\ref{specify-semilocal-sum} with $O(n^2)$ polynomials of bounded degree in $O(n)$ variables.
The information contained in the specification can be described by $n^{O(1)}$
conditions, in the forms of Eqns~(\ref{Gij}) and (\ref{gFF1},\ref{gFF2}), in $n^{O(1)}$ unknowns representing the coefficients of the hidden polynomials, including  $F_{ij}$, which determine a basic blinding map, and $g_{ij}$, $g'_{ij}$, which determine the local functions, with $i=1,\ldots,n$, $j=1,2$.
Let $V_{\varphi}$ be the algebraic set determined by these conditions.
Let $V_{\la\varphi\ra}$ be the algebraic set determined by these conditions, however
with $F_{ij}$ expressed in terms of the blinding parameters.
Let $\alpha$ be a point of  $V_{\varphi}$ (resp. $V_{\la\varphi\ra}$).  Then $\alpha$ determines, through a subset of its coordinates, a set $S$ of $9n^2$ semi-local functions of the same type as those involved in specifying $\hat{\varphi}$, and $V_S$ (resp. $V_{\la S\ra}$) is locally embedded around $\alpha$.
\end{theorem}

\begin{theorem}
\label{specify-S}
Suppose a random basic blinding map $\rho$ is chosen and
 a set $\cF$ of functions each of which is either the sum or product of $O(n)$ semi-local functions of bounded degree is specified.
Then
the information contained in the specification can be described by $n^{O(1)}$
conditions, in the forms of Eqns~(\ref{lij-sum},\ref{lij-product}),(\ref{Gij}) and (\ref{gFF1},\ref{gFF2}), in $n^{O(1)}$ unknowns representing the coefficients of the hidden polynomials, including  $F_{ij}$, with $i=1,\ldots,n$ and $j=1,2$, which determine a basic blinding map, and polynomials which determine the hidden local functions.
Let $V_{\cF}$ be the algebraic set determined by these conditions.
Let $V_{\la \cF \ra}$ be the algebraic set determined by these conditions, however with $F_{ij}$ expressed in terms of the blinding parameters.
Let $\alpha$ be a point of  $V_{\cF}$ (resp. $V_{\la\cF\ra}$).  Then $\alpha$ determines, through a subset of its coordinates, a set $S$ of semi-local functions of the same type as those involved in  $\cF$, and $V_S$ (resp. $V_{\la S\ra}$) is locally embedded around $\alpha$.
\end{theorem}

\ \\Theorem~\ref{specify-S} together with Theorem~\ref{semi-local-thm} underscores the difficulty of solving for a closed point of $V_{\cF}$ or $V_{\la \cF \ra}$ for the purpose of un-blinding.
\subsection{Blinding maps with Frobenius twists}
\label{Frob}
Fix and publish a randomly chosen basis of $K/k$, $\bt = \theta_1,\ldots,\theta_d$.
As before, let $\tau$ denote the Frobenius map $x\to x^q$ for $x\in\bar{k}$ (where $k=\F_q$), let $\tau_a = \tau^a$, and $\tau_{a,b}$ denote the Frobenous twist $\bar{k}^2\to\bar{k}^2: (x,y)\to (\tau_a(x), \tau_b(y))=(x^{q^a},y^{q^b})$.
When the context is clear we also denote $\tau_{a,a}$ as $\tau_a$.

\ \\We consider a blinding map $\rho'=(\rho'_i)_{i=1}^n$ of the form $\rho'_i=(\tau_{a_i, b_i}\circ\rho_i)_{i=1}^n$ where $\rho=(\rho_i)_{i=1}^n$ is a basic blinding map.  Suppose $\rho_i =(F_{i1},F_{i2})$ where $F_{ij}$ are quadratic polynomials in $3n$ variables.

\ \\For diagonal matrix $A=\left( \begin{array}{cc} \alpha & 0 \\ 0 & \beta \end{array}\right)$, let $A^{\tau_{a,b}} =
 \left( \begin{array}{cc} \alpha^{\tau_a} & 0 \\ 0 & \beta^{\tau_b} \end{array}\right)$.  It is easy to verify that
 $\tau_{a,b}\circ A = A^{\tau_{a,b}} \circ \tau_{a,b}$ as maps on $\bar{k}^2$. Now suppose $g$ has semi-local decomposition $[f]\circ [\rho']$ where $f$ is a local function. In this case we may write the decomposition as $[f]\circ [\tau_{{\bf a}, {\bf b}}]\circ [\rho]$, where
$\tau_{{\bf a},{\bf b}} = (\tau_{a_1,b_1},\ldots,\tau_{a_n,b_n})$. Let $A$ be a diagonal matrix with $A_1$, ... $A_n$ as the diagonal blocks where $A_i \in Gl_2 (\bar{k})$ is diagonal.   Let $A^{\tau_{{\bf a},{\bf b}}}$ denote the diagonal matrix with $A_1^{\tau_{a_1,b_1}}$, ... $A_n^{\tau_{a_n,b_n}}$ as the diagonal blocks.
Then $g$ has semi-local decomposition $[f \circ A^{\tau_{{\bf a},{\bf b}}}] \circ [\tau_{{\bf a},{\bf b}}]\circ [A^{-1}\rho]$.
 From this it is not hard to see the that we have the following generalization of Theorem~\ref{semi-local-thm}.

\begin{theorem}
\label{semi-local-thm-frob-twist}
Suppose we have a set of semi-local functions $g_i$, $i=1,\ldots,m$, such that $g_i$ has semi-local decomposition
$[f_i]\circ [\rho']$ for all $i$, where $f_i$ is a local function and $\rho'=\tau_{{\bf a},{\bf b}}\circ\rho$ where $\rho$ is a basic blinding map and $\tau_{{\bf a},{\bf b} }= (\tau_{a_1,b_1},\ldots,\tau_{a_n,b_n})$.
Then the following hold.
\begin{enumerate}
\item
There is an injective map $\bar{k}^{2n^2}\to [\rho]$, and an injective map  $\bar{k}^{n}\to\la \rho\ra$.
\item
There is an injective map $\bar{k}\to [f_i]$ if $f_i$ is $c$-local depending on $V_{i_1}\times\ldots V_{i_c}$ and for some $j$ the degree of $f_i$ at $x_{i_j}$ is greater to equal to the minimum degree of polynomials in the ideal defining $V_{i_j}$.
\item
Let $A_j \in Gl_2 (\bar{k})$ be a diagonal matrix for $j=1,\ldots,n$.  Let $A$ be the block-diagonal matrix with $A_1$, ..., $A_n$ as the diagonal blocks.  Then $g_i$ has semi-local decomposition
 $[f_i \circ A^{\tau_{{\bf a},{\bf b}}}] \circ [\tau_{{\bf a},{\bf b}}]\circ [A^{-1}\rho]$
 for $i=1,\ldots,m$.
\end{enumerate}
\end{theorem}

\ \\Let $S=\{g_i: i=1,\ldots,m\}$ be as in the theorem.  Let $V_S$ be the union of $[f_1]\times\ldots\times [f_m]\times [\rho]$, where the union is over all $f_1,\ldots,f_m,\rho$ such that $g_i$ has semi-local decomposition $[f_i]\circ [\tau_{{\bf a},{\bf b}}] \circ [\rho]$, $i=1,\ldots,m$.  Similarly let $V_{\la S\ra}$ be the union of $[f_1]\times\ldots\times [f_m]\times \la\rho\ra$, where the union is over all $f_1,\ldots,f_m,\rho$ such that $g_i$ has semi-local decomposition $[f_i]\circ [\tau_{{\bf a},{\bf b}}] \circ \la\rho\ra$, $i=1,\ldots,m$.  Then $V_S$ and $V_{\la S\ra}$ admit local embedding of affine space of dimension $\Omega(n^2)$ (respectively $\Omega(n)$) around every point by the first two assertions of Theorem~\ref{semi-local-thm-frob-twist}, and both are acted on by the subgroup of diagonal matrices of $Gl_{2n} (\bar{k})$ in a twisted fashion.
The two properties combined, and the fact that their dimensions are huge (respectively $\Omega(n^2)$ and $\Omega(n)$), seem to make it difficult to solve for a closed point even if polynomial systems describing $V_S$ and $V_{\la S\ra}$ are known.

\ \\Suppose $f$ is a polynomial of degree $d$ then $\tau_{a} \circ f (\bx) = (f (\bx))^{q^a}$ has degree $dq^a$.
Therefore polynomials of degree exponential in $q$ result as we apply Frobenius twists to blind a semi-local function.
This makes it more complicated to describe $V_S$ and $V_{\la S\ra}$.

\ \\In order to specify blinded maps using low degree polynomials we consider the {\em descent} trick which is involved in Weil restriction (descent).

\ \\Suppose $F\in \bar{k}[x_1,\ldots,x_{3n}]$.  Let $\tilde{x}_i = \sum_{j=1}^d x_{ij} \theta_j$, for $i=1,\ldots,3n$, where $x_{ij}$ are variables.  Let $\bx = x_1,\ldots,x_{3n}$ and $\hat{\bx} = x_{11},\ldots,x_{3n,d}$.  Let $\tilde{F} (\hat{\bx}) =   F (\tilde{x}_1,\ldots, \tilde{x}_{3n})$. We call $\tilde{F}$ the {\em descent} of $F$ with respect to $\bt$, or simply the descent of $F$ when $\bt$ is fixed.

\ \\Let $J$ be the ideal generated by $x_{ij}^q -x_{ij}$ for all $i,j$.  Let $\tilde{F} \mod J$ denote the polynomial $G(\hat{\bx})$ with degree less than $q$ in all $x_{ij}$ such that $\tilde{F}(\hat{\bx})\equiv G(\hat{\bx}) \mod J$.
For $\alpha_1,\ldots,\alpha_{3n}\in K$,  let $\hat{\alpha}_i\in k^d$ such that $\alpha_i =\la \hat{\alpha}_i , \bt\ra$ for all $i$.  Suppose $G = \tilde{F}\mod J$.  Then $G(\hat{\alpha}_1,\ldots, \hat{\alpha}_{3n}) = \tilde{F} (\hat{\alpha}_1,\ldots, \hat{\alpha}_{3n}) =F (\alpha_1,\ldots,\alpha_{3n})$.

\ \\Note that  $\widetilde{x_i^{q^a}} \equiv \sum_{j=1}^d x_{ij} \theta_j^{\tau_a} \mod J$.  So for $F\in \bar{k}[x_1,\ldots,x_{3n}]$,
$$\widetilde{ F( x_1^{q^{a_1}},\ldots, x_{3n}^{q^{a_{3n}}})}\equiv F(\ldots, \sum_{j=1}^d x_{ij} \theta_j^{\tau_{a_i}},\ldots)  \mod J.$$
Therefore $\widetilde{ F( x_1^{q^{a_1}},\ldots, x_{3n}^{q^{a_{3n}}})}\mod J$
has degree bounded in $\deg F$.

\ \\To specify a semi-local function $g=f\circ\tau_{{\bf a}, {\bf b}}\circ \rho$ which is applied to $K$-points, it suffices to specify $\tilde {g} \mod J$, which, from the discussion above, is of degree bounded in the degree of $f\circ\rho$, which in our context is $O(1)$.

\ \\We have an injective homomorphism $\bar{k}[\bx]\to\bar{k}[\hat{\bx}]: f\to\tilde{f}$.   Let $\tilde{V}_{S}$ (resp. $\tilde{V}_{\la S\ra}$) denote the image of $V_S$ (resp. $V_{\la S\ra}$) under the map naturally induced by the injection $[\rho]\to [\tilde{\rho}]$.   From this observation and Theorem~\ref{semi-local-thm-frob-twist} we have the following theorem.

\begin{theorem}
\label{semi-local-thm-frob-twist-tilde}
Suppose we have a set $S$ of semi-local functions $g_i$, $i=1,\ldots,m$, such that $g_i$ has semi-local decomposition
$[f_i]\circ [\rho']$ for all $i$, where $f_i$ is a local function and $\rho'=\tau_{{\bf a},{\bf b}}\circ\rho$ where $\rho$ is a basic blinding map and $\tau_{{\bf a},{\bf b} }= (\tau_{a_1,b_1},\ldots,\tau_{a_n,b_n})$.
Then the following hold.
\begin{enumerate}
\item
Around every point of $\tilde{V}_S$ there is an embedding of $\bar{k}^{2n^2}$ relative to the blinding part, and an embedding of $\bar{k}$ relative to the local part if there is some $f_i$, $c$-local depending on $V_{i_1}\times\ldots V_{i_c}$, where for some $j$ the degree of $f_i$ at $x_{i_j}$ is greater to equal to the minimum degree of polynomials in the ideal defining $V_{i_j}$.
\item
Around every point of $\tilde{V}_{\la S\ra}$ there is an embedding of $\bar{k}^{n}$ relative to the blinding part, and an embedding of $\bar{k}$ relative to the local part if there is some $f_i$, $c$-local depending on $V_{i_1}\times\ldots V_{i_c}$, where for some $j$ the degree of $f_i$ at $x_{i_j}$ is greater to equal to the minimum degree of polynomials in the ideal defining $V_{i_j}$.

\item
There is a twisted action of the subgroup of diagonal matrices of $Gl_{2n} (\bar{k})$ on $\tilde{V}_S$ and $\tilde{V}_{\la S\ra}$.
\end{enumerate}
\end{theorem}

\ \\For specifying a function that is the sum of semi-local functions, the same procedure in \S~\ref{specify-semilocal-sum} applies.  Then the specifying polynomials $g''_i$ and $h''_i$ are now to be expressed in descent form by applying the substitution
$x_i = \sum_{j=1}^d x_{ij}\theta_j$
for $i=1,\ldots,3n$.  The conditions (\ref{lij-sum}) become in this setting the following:
\begin{eqnarray}
\label{lij-sum-tilde}
\frac{h_{i1}\circ(\tau_{a_{i_1}, b_{i_1}}, \tau_{a_{i_2},b_{i_2}}) \circ \tilde{\bf F}_i \mod J }{h_{i2}\circ(\tau_{a_{i_1},b_{i_1}}, \tau_{a_{i_2},b_{i_2}}) \circ \tilde{\bf F}_i \mod J } +\frac{\tilde{\ell}_{i,1}}{\tilde{\ell}_{i,2}} - \frac{\tilde{\ell}_{i+1,1}}{\tilde{\ell}_{i+1,2}} =_{\tilde{W}} \frac{\tilde{g}''_i}{\tilde{h}''_i}.
\end{eqnarray}
\ \\A similar analysis can be carried out for products of semi-local functions, maps of bounded locality.
Proceeding in a way similar to the proof of Theorem~\ref{semi-local-product-thm}, Theorem~\ref{specify-hat} and Theorem~\ref{specify-S},  we obtain analogous theorems for blinding with Frobenius twists.  We state below the analogous theorem to Theorem~\ref{specify-S}.
\begin{theorem}
\label{specify-S-tilde}
Suppose a random blinding map $\rho$ is chosen and
 a set $\cF$ of functions each of which is either the sum or product of $O(n)$ semi-local functions of bounded degree is specified as discussed in this section.
Then we have the following.
\begin{enumerate}
\item
The information contained in the specification can be described by $n^{O(1)}$
conditions, in $n^{O(1)}$ unknowns representing the hidden polynomials in the descent form, including  $\tilde{F}_{ij}$, with $i=1,\ldots,n$ and $j=1,2$, where $F_{ij}$'s determine a basic blinding map, and the polynomials which determine the hidden local functions.
\item
Let $\tilde{V}_{\cF}$ be the algebraic set determined by these conditions.
Let $\tilde{V}_{\la \cF \ra}$ be the algebraic set determined by these conditions, however with $\tilde{F}_{ij}$ expressed in terms of the blinding parameters.
Let $\alpha$ be a point of  $\tilde{V}_{\cF}$ (resp. $\tilde{V}_{\la\cF\ra}$).  Then $\alpha$ determines, through a subset of its coordinates, a set $S$ of semi-local functions of the same types and degrees as those involved in  $\cF$, and $\tilde{V}_S$ (resp. $\tilde{V}_{\la S\ra}$) is locally embedded around $\alpha$.
\end{enumerate}
\end{theorem}

\ \\Suppose $\cF$ contains a constant number of functions and each function is the sum or product of a constant number of semi-local functions.  Then finding a point on $\tilde{V}_{\cF}$ or $\tilde{V}_{\la \cF \ra}$ reveals a blinding partially at the localities involved in the semi-local functions.  Even in this case the number of unknown in describing $\tilde{V}_{\cF}$ is $\Omega(n^3)$ and the number of unknown describing $\tilde{V}_{\la \cF \ra}$ is $\Omega(n^2)$, and both algebraic sets are of dimension $\Omega(n)$.  Applying best known methods to find a point on  $\tilde{V}_{\cF}$ (resp. $\tilde{V}_{\la \cF \ra}$) takes time exponential in  $O(n^3\log n)$ (resp. $O(n^2\log n )$) (\cite{Ay,HW}). Therefore Theorem~\ref{specify-S-tilde}  serves as a strong evidence that the blinding is secure.
Moreover the triply confusing property stated in Theorem~\ref{semi-local-thm-frob-twist-tilde} provides additional evidence that efficient algorithms to find points on such algebraic sets seems unlikely to exist.
\subsection{Application to trilinear map}
Recall that to construct a trilinear map we start by choosing from the isogeny class of a pairing friendly elliptic curve some $E/K$ defined by $y^2=x^3+ax+b$ with $a,b\in K$.  We assume that $E[\ell] \subset E(K)$, where $\log\ell$ and $\log |K|$ are linear in the security parameter.  The curve $E$ is considered secret, as well as the set $\{ \overline{E^A}: A\in Gl_2 (K)\}$.  Recall that
for $A\in Gl_2 (K)$, $E^A$ denote the elliptic curve which is the image of $E$ under $A$, and $\overline{E^A}$ denote the image of $E^A$ under $\jmath:(x,y)\to (x^{-1}, y^{-1})$. Choose randomly from this family $E_i$, $i=1,\ldots,n$.

\ \\As in \S~\ref{Frob}, suppose $K$ is a finite extension over $k=\F_q$.  Fix and publish a randomly chosen basis of $K/k$, and let $\tau$ denote the Frobenius map $x\to x^q$ for $x\in\bar{k}$.
Choose a random blinding map $\rho=(\rho_i)_{i=1}^n$ of the form $\rho_i=(\tau_{a_i}\circ\rho'_i)_{i=1}^n$ where $\rho'=(\rho'_i)_{i=1}^n$ is a basic blinding map.
Let $\hat{E}=\rho^{-1} \prod_{i=1}^n E_i$.
Choose $\alpha,\beta\in E[\ell]^n$ such that $e(\alpha,\beta)\neq 1$.  Let $\hat{\alpha},\hat{\beta}\in \hat{E}[\ell]$ such that $\hat{\alpha}$ corresponds to $\alpha$ and $\hat{\beta}$ corresponds to $\beta$ under
$\hat{E}\stackrel{\rho}{\to} \prod_{i=1}^n E_i \simeq E^n$.

\ \\The trilinear map is specified to the public by
$\hat{\alpha}, \hat{\beta}\in \hat{E}[\ell]\subset K^{3n}$,
the addition morphism $\hat{m}$ (with the doubling map separately specified as a subcase), $\hat{\varphi}_i$, $i=1,\ldots, N$ where $N=O(n^2)$; and for the computation of $\hat{e}$ two functions $\hat{g}$ and
$\hat{h}$ are specified, both are products of semi-local functions of bounded degree, as will be discussed in \S~\ref{pairing}.  The addition map on $\hat{E}[\ell]$, $\hat{m}$, can be securely specified by generalization of Theorem~\ref{specify-hat} (as discussed in \S~\ref{Frob} see also Theorem~\ref{specify-S-tilde}).
Theorem~\ref{specify-hat}  can be applied to specify $\hat{\varphi}_i$.  Theorem~\ref{specify-S-tilde} can be applied to specify $\hat{g}$ and $\hat{h}$.

\begin{theorem}
\label{specify-trilinear}
The information contained in the specification of the trilinear map described in this section can be described by $n^{O(1)}$ algebraic
conditions in $m$ unknown where $m=n^{O(1)}$ and $m=\Omega(n^2)$.
Let $\tilde{V}_{{\cal T}}$ be the algebraic set determined by these conditions.
Let $\tilde{V}_{\la {\cal T} \ra}$ be the algebraic set determined by these conditions, however with the quadratic polynomials describing the basic blinding map that is  involved expressed in terms of the blinding parameters.
Let $S_0$ be the set of hidden semi-local functions that are involved in specifying the trilinear map.
Then $\tilde{V}_{S_0}$ (resp. $\tilde{V}_{\la S_0\ra}$) is triply confusing of dimension $\Omega(n^2)$ (resp. $\Omega(n)$).  For every point of $\tilde{V}$ (resp. $\tilde{V}_{\rho}$) either $\tilde{V}_{S_0}$ (resp. $\tilde{V}_{\la S_0\ra}$) or some triply confusing $\tilde{V}_{S}$ (resp. $\tilde{V}_{\la S\ra}$) can be embedded around the point, where $S$ is a set of semi-local functions of the same type as $S_0$.
\end{theorem}

\ \\{\bf Proof} Let $S_0$ be the set of semi-local functions involved in $\hat{m}$, $\hat{\varphi_i}$, $\hat{g}$ and $\hat{h}$.  The local functions involved in these semi-local functions are all related to the addition law on $E$ and are of degree 3 in at least one variable (see \S~\ref{pairing}).  This implies by Theorem~\ref{semi-local-thm} the algebraic set $V_{S_0}$ and $V_{\la S_0\ra}$ are triply confusing, so are $\tilde{V}_{S_0}$ and $\tilde{V}_{\la S_0\ra}$.
Theorem~\ref{specify-trilinear} follows as Theorem~\ref{specify-S} and Theorem~\ref{specify-S-tilde} are applied to the current context.  $\Box$

\ \\The remark below Theorem~\ref{specify-S-tilde} can be similarly made.  In addition the algebraic sets described in the theorem are most likely uniformly confusing by virtue of the choice of $E_i$ from the set $\{ \overline{E^A} : A\in Gl_2 (K)\}$.  Therefore Theorem~\ref{specify-trilinear}  serves as a strong evidence that the blinding is secure.

\section{Pairing computation}
\label{pairing}
To complete the description of the trilinear map we demonstrate in this section how $\hat{e}$ can be explicitly defined and specified, and efficiently computed.
To simply notation and make the presentation easier we identify $E_i$ and $E$ through isomorphism and denote for example the addition morphism of $E_i\simeq E$ also as $m$.  We also assume the same blinding map $\rho$ is used to form the pairing groups, so that $\hat{e}$ is a pairing on $\hat{E}[\ell]$.

\ \\Suppose the characteristic of $K$ is not 2 or 3, and $E$ is given $y^2 =x^3 +ax+b$ with $a,b\in K$.
The addition map of $E$ can be described as follows (see \cite{Sil}).
Let $P_1 = (x_1,y_1)$, $P_2 = (x_2,y_2)$ be two points on $E$.
If $x_1 = x_2$ and $y_1 = -y_2$, then $P_1 + P_2 = 0$.
Otherwise, we can find $P_3 = (x_3,-y_3)$ such that
$P_1$, $P_2$ and $\bar{P}_3 = (x_3,y_3)$ lie on a line
$y = \lambda x + \nu$, and we have
$P_1 + P_2 = P_3$.

\ \\(1) If $x_1 \neq x_2$, then $\lambda = \frac{y_2-y_1}{x_2-x_1}$ and $\nu = \frac{y_1 x_2 - y_2 x_1}{x_2-x_1}$.

\ \\(2) If $x_1 =  x_2$ and $y_1 \neq 0$, then $\lambda = \frac{3x_1^2 +a}{2y_1}$ and $\nu = \frac{- x^3_1+ax_1+2b}{2y_1}$

\ \\In both cases
$x_3 = \lambda^2 -x_1-x_2$, $y_3 = -\lambda x_3 -\nu$.

\ \\Suppose $P_i = (x_i,y_i)$ with $P_i\in E$ for $i=1,2,3$, $x_1\neq x_2$ and $P_1 +P_2 = P_3$.
Then $g_{P_1,P_2}:=\frac{y-\lambda x+\nu}{x-x_3}+P_3 -O = P_1+P_2 -2O$, where $\lambda = \frac{y_2-y_1}{x_2-x_1}$ and $\nu = \frac{y_1 x_2 - y_2 x_1}{x_2-x_1}$.  Let $g: E\times E\times E \to \bar{k}$ such that for $P_1=(x_1,y_1),P_2=(x_2,y_2)\in E$, and $Q=(x,y)\in E$,
$g(P_1,P_2,Q) = \frac{y-\lambda x+\nu}{x-x_3} = g_{P_1,P_2} (Q)$.

\ \\Let $\hat{g}:\hat{E}\times\hat{E}\times\hat{E} \to \bar{k}$ so that for $\hat{\alpha},\hat{\beta},\hat{\gamma}\in\hat{E}$.
$\hat{g} (\hat{\alpha},\hat{\beta},\hat{\gamma})= \prod_{i=1}^n g (\rho_i (\hat{\alpha}),\rho_i (\hat{\beta}), \rho_i (\hat{\gamma}) )$.

\ \\For $D=P_1 - O$ where $P_1 = (x_1,y_1)$ is not 2-torsion, we have $2D = (h_D) + D'$ where $D' = P_3-O$ with
$2P_1 = P_3 = (x_3,y_3)$ given by the formula above, and $h_D (x,y)  = \frac{L}{x-x_1}$ where $L=y-\lambda x -\nu$,
$\lambda = \frac{3x_1^2 +a}{2y_1}$ and $\nu = \frac{- x^3_1+ax_1+2b}{2y_1}$.  So let $h: E\times E \to \bar{k}$ so that for
$P_1 = (x,y)\in E$ and $P_2 = (x_2,y_2)\in E$, $h(P_1,P_2) = h_D (P_2)=h_D (x_2,y_2)$ as above where $D=P_1-O$.

\ \\Let $\hat{h}: \hat{E}\times\hat{E} \to \bar{k}$ so that for $\hat{\alpha},\hat{\beta}\in\hat{E}$.
$\hat{h} (\hat{\alpha},\hat{\beta})= \prod_{i=1}^n h (\rho_i (\hat{\alpha}),\rho_i (\hat{\beta}))$.

\ \\Suppose $P\in E[\ell]$.  Then $D=P-O$ is an $\ell$-torsion divisor.  We recall how to efficiently construct $h$ such that $\ell D = (h)$ through the squaring trick \cite{Miller,Miller1}.

\ \\Let $D_i = P_i -O$ where $P_i = 2^i P$ for all $i$. Apply addition to double $D$, and get
\[ 2D = (h_D) + D_1 .\]
Inductively, we have $H_i$ such that
\[ 2^i D = (H_i) + D_i .\]
Apply addition to double $D_i$ and get
\[ 2D_i = (h_{D_i}) + D_{i+1}.\]   We have
\[ 2^{i+1} D = (H_{D,i+1}) + D_{i+1}\]
where $H_{D,i+1} = H_{D,i}^2 h_{D_i}$.

\ \\Write $\ell = \sum_i a_i 2^i$ with $a_i\in\{0,1\}$.
Let $H_D = \prod_i H_{D,i}^{a_i}$.  Then
$\ell D = (H_D)+\sum_i a_i D_{i+1}$.

\ \\Write $\sum_i a_i D_{i+1} = D_{i_1}+\ldots+D_{i_m}$ with $i_1 >\ldots > i_m$.
 Then $P_{i_1}+P_{i_2} = Q_2$, $Q_2 + P_{i_3} = Q_3$, ..., $Q_{i_{m-1}} + P_{i_m} = O$ with
 $Q_j\in E$ for $j=1,\ldots,m-1$.  We have $\sum_i a_i D_{i+1} = (G_D)$ where
 $G_D = g_{P_{i_1},P_{i_2}} g_{Q_2,P_{i_3}}\ldots g_{Q_{i_{m-1}}, P_{i_m}}$.

 \ \\We have $\ell D = (H_D G_D)$. Let $f_{P}= H_D G_D$.  Then for $P,Q\in E[\ell]$,
 $e_E (P,Q) = \frac{f_P (Q)}{f_Q (P)}$, where $e_E$ is the Weil pairing on $E[\ell]$.

\ \\Suppose $\hat{\alpha}\in\hat{E}[\ell]$.  Let $\hat{\alpha}_i = 2^i \hat{\alpha}$.  Then for $j=1,\ldots,n$,
$2 \rho_j \hat{\alpha}_i = \rho_j \hat{\alpha}_{i+1}$.

\ \\Let $\hat{D}=\hat{\alpha}-O$.  Let $h$ be as defined before where  $h(P_1,P_2) = h_{P_1-O} (P_2)$ for $P_1,P_2\in E$.
Inductively define $\hat{H}_{i+1} =\hat{H}_i^2 \hat{h}$.
We can verify inductively
\[\hat{H}_{j} (\hat{\alpha},\hat{\beta})= \prod_{i=1}^n H_{\rho_i(\hat{D}), j} ( \rho_i ( \hat{\alpha}), \rho_i (\hat{\beta}))\]

\ \\Let $\hat{H} = \prod_i \hat{H}_i^{a_i}$.  Then $\hat{H} (\hat{\alpha},\hat{\beta}) = \prod_{i=1}^n H_{\rho_i (\hat{\alpha})-O} (\hat{\beta})$, and can be efficiently computed once $\hat{h}$ is specified.

\ \\Write $\sum_i a_i \hat{\alpha}_{i+1} = \hat{\alpha}_{i_1}+\ldots+\hat{\alpha}_{i_m} =O$ with $i_1 > \ldots >i_m$.
Let $\hat{\beta}_i$ be such that $\hat{\alpha}_{i_1}+\hat{\alpha}_{i_2} = \hat{\beta}_2$,
$\hat{\beta}_2 +\hat{\alpha}_{i_3} = \hat{\beta}_3$,$\ldots$, $\hat{\beta}_{i_{m-1}} + \hat{\alpha}_{i_m} = O$.
We have
\begin{eqnarray*}
 \hat{g} (\hat{\alpha}_{i_1},\hat{\alpha}_{i_2},\hat{\beta}) & = & \prod_{i=1}^n g (\rho_i (\hat{\alpha_{i_1}}), \rho_i (\hat{\alpha}_{\i_2}),\rho_i (\hat{\beta}))\\
 \hat{g} (\hat{\beta}_{2},\hat{\alpha}_{i_3},\hat{\beta}) & = & \prod_{i=1}^n g (\rho_i (\hat{\beta_{2}}), \rho_i (\hat{\alpha}_{\i_3}),\rho_i (\hat{\beta}))\\
 \ldots \\
 \hat{g} (\hat{\beta}_{m-1},\hat{\alpha}_{i_m},\hat{\beta}) & = & \prod_{i=1}^n g (\rho_i (\hat{\beta_{m-1}}), \rho_i (\hat{\alpha}_{\i_m}),\rho_i (\hat{\beta}))
\end{eqnarray*}
So
\[ \hat{g} (\hat{\alpha}_{i_1},\hat{\alpha}_{i_2},\hat{\beta}) \hat{g} (\hat{\beta}_{2},\hat{\alpha}_{i_3},\hat{\beta})
\ldots \hat{g} (\hat{\beta}_{m-1},\hat{\alpha}_{i_m},\hat{\beta}) = \prod_{i=1}^n G_{\rho_i (\hat{\alpha}-O )} (\rho_i (\hat{\beta}))
\]
Therefore $\prod_{i=1}^n f_{\rho_i (\hat{\alpha})} (\rho_i (\hat{\beta}))$ can be computed efficiently using $\hat{g}$ and $\hat{h}$.

\ \\Similaly $\prod_{i=1}^n f_{\rho_i (\hat{\beta})} (\rho_i (\hat{\alpha}))$ can be computed efficiently using $\hat{g}$ and $\hat{h}$.
So $\hat{e} (\hat{\alpha},\hat{\beta}) = \frac{\prod_{i=1}^n f_{\rho_i (\hat{\alpha})} (\rho_i (\hat{\beta}))}{\prod_{i=1}^n f_{\rho_i (\hat{\beta})} (\rho_i (\hat{\alpha}))}$ can be computed efficiently using $\hat{g}$ and $\hat{h}$.

\ \\Finally we note that both $\hat{g}$ and $\hat{h}$ are products of semi-local functions.  They can be specified securely using the procedure described in \S~\ref{specify-semilocal-product}.

\section*{Acknowledgements}
I would like to thank the participants of  the AIM workshop on cryptographic multilinear maps (2017), and the participants of the BIRS workshop: An algebraic approach to multilinear maps for cryptography (May 2018), for stimulating and helpful discussions. I would especially like to acknowledge the contributions of the following colleagues:
Dan Boneh and Amit Sahai for valuable discussions during the early phase of this work;
Steven Galbraith for careful reading of the preprint in \cite{H1} as well as valuable comments and questions;
Steven Galbraith, Karl Rubin, Travis Scholl, Shahed Sharif, Alice Silverberg, and Ben Smith for valuable comments and questions on a subsequent preprint \cite{H2}.

\end{document}